\documentclass[psfig]{aa} 

\usepackage{times}
\usepackage{graphics}
\usepackage{latexsym}
\usepackage{epsfig}

\begin{document}
\title{
The ISOPHOT\,-\,MAMBO survey of 3CR radio sources\thanks{Based on
  observations with the Infrared Space Observatory 
  ISO, an ESA project with instruments 
  funded by ESA Member States (especially the PI countries: France, 
  Germany, the Netherlands and the United Kingdom) and
  with the participation of ISAS and NASA.}:\\
Further evidence for the unified schemes
}
\author{M.\ Haas\inst{1,2}
   \and S.A.H.\ M\"uller\inst{2}
   \and F.\ Bertoldi \inst{3}
   \and R.\ Chini\inst{2}
   \and S.\ Egner\inst{1}   
   \and W.\ Freudling\inst{4}
   \and U.\ Klaas\inst{1}
   \and \\
        O.\ Krause\inst{1} 
   \and D.\ Lemke\inst{1} 
   \and K.\ Meisenheimer\inst{1}
   \and R.\ Siebenmorgen\inst{4}
   \and I.\ van Bemmel\inst{5}
}
\offprints{Martin Haas (haas@astro.rub.de)}
\institute{
Max--Planck--Institut f\"ur Astronomie (MPIA), K\"onigstuhl 17, D-69117
Heidelberg, Germany
\and
Astronomisches Institut, Ruhr-Universit\"at Bochum (AIRUB),
Universit\"atsstr. 150 / NA7, D-44780 Bochum, Germany
\and
Max--Planck--Institut f\"ur Radioastronomie (MPIfR), Auf dem H\"ugel 69, D-53121 
Bonn, Germany
\and
European Southern Observatory (ESO), Karl-Schwarzschild-Str. 2, D-85748 Garching, Germany
\and
Space Telescope Science Institute (STScI), 3700 San Martin Drive, Baltimore, MD 21218, USA
}
\date{received 1. December 2003 ; accepted 3. June 2004}
\authorrunning{M. Haas et al.}
\titlerunning{3CR radio galaxies and quasars}

\abstract{
We present the complete set of ISOPHOT observations of 3CR radio galaxies and quasars,
which are contained in the ISO Data Archive, 
providing 75 mid- and far-infrared spectral energy distributions
(SEDs) between 5 and 200\,$\mu$m. For 28 sources they are   
supplemented with MAMBO 1.2 mm observations and for
15 sources with new submillimetre
data from the SCUBA archive. 
The sample includes flat and steep spectrum quasars, broad
and narrow line radio galaxies, as well as
Fanaroff-Riley FR\,1 and FR\,2 types.
The SED shapes exhibit a diversity in the infrared (IR),
ranging from a smooth dominating synchrotron 
component in flat spectrum sources to a thermal dust bump  around 60-100 $\mu$m 
in steep spectrum sources.
The  detection rate of a thermal  bump in more than 50\% of
the cases suggests that dust emission may be a general phenomenon in
these sources.
We check the orientation-dependent unified scheme, in which 
the powerful FR\,2 narrow line galaxies are quasars viewed at high inclination, so that
their nuclei are hidden behind a dust torus
intercepting the optical-ultraviolet AGN radiation and reemitting it in the infrared.
We find that (1) 
both the quasars and the galaxies show a high mid- to
far-infrared luminosity ratio typical for powerful AGNs  and (2)
-- when matched in 178 MHz luminosity -- both 
show the same ratio of isotropic far-infrared to isotropic 178
MHz lobe power.
Therefore, from our large sample investigated here
we find strong evidence for the orientation-dependent unification of the powerful
FR\,2 galaxies with the quasars.
The distribution of the dust-to-lobe luminosity ratio shows a dispersion which
we suggest to be most likely due to the
additional influence of evolution and environment superposed on the
orientation-dependent unification.
We discuss our data also in the frame of the receding torus model. 
At the high 178 MHz luminosities of our sources above 10$^{\rm 26.5}$ W/Hz
we do not find any support for this model in its original formulation
and therefore we propose a refinement:
The scale height of the torus might not be independent of luminosity,
rather it may increase 
at high luminosities due to the impact of supernovae from starbursts accompanying the AGN phenomena.


}
\maketitle

\section{Introduction} 
\label{section_introduction}

\subsection{Unified schemes} 
\label{section_unified_schemes}

While observationally at least a dozen classes of active 
galactic nuclei (AGN) are discernible, 
on the conceptional side consensus is growing that in most, 
if not all, of these different objects 
the accretion onto a massive 
black hole (M$_{\rm bh}$ $\ge$ 10$^{\rm 6}$ M$_{\odot}$) provides the main source of 
energy. It seems natural to explain their apparent 
differences by observational  effects such as (1) the {\it circumgalactic environment}, (2) 
the {\it evolutionary phase}, or (3) the {\it aspect angle}. 
To disentangle these effects is one of the major challenges of the current AGN research, as 
reviewed by e.g. Urry \& Padovani (1995).

The Revised Third Cambridge Catalogue of radio galaxies and quasars
at 178 MHz
by Laing et al. (1983, abbreviated 3CRR) and -- applying softer criteria -- 
by Spinrad et al. (1985, for short 3CR) contains 178 and 298 sources,
respectively, which fall into several classes of radio-loud AGN:
\begin{itemize}
  \item[1)] flat spectrum quasars, which are almost core dominated,
  \item[2)] steep spectrum  broad line radio
  galaxies (BLRGs) and quasars (here collectively called {\it quasars}), with either compact or
  lobe dominated morphology,
  \item[3)] powerful narrow line radio galaxies (NLRGs) of
  Fanaroff-Riley type FR\,2 with -- often asymmetric -- jets ending in 
  edge-brightened double lobes extending far beyond the host galaxy, 
  \item[4)] low power narrow line radio galaxies of
  Fanaroff-Riley type  FR\,1, with double jets feeding edge-darkened 
    radio lobes often contained within the host galaxy.
\end{itemize}
The diversity and the brightness of the sources makes the 3CR catalogue
a well suited database for studying the relationship between these AGN
classes.
The aspect angle relation between flat and steep spectrum quasars,
with 
the flat spectrum sources being steep spectrum sources seen pole-on
(as proposed by Orr \& Browne 1982), is
now widely accepted. The relation between steep spectrum quasars and
FR\,2 radio galaxies and that between FR\,1 and FR\,2 sources is still
a matter of debate.

While the quasars show the optical signatures of a powerful AGN 
(like broad emission lines and high excitation and optical 
luminosity), the galaxies do not.
In an early attempt, physically different central engines were
proposed as reviewed by Begelman, Blandford \&
Rees (1984): quasars have a high mass accretion rate but a low black hole spin,
and galaxies have a low  mass accretion rate but a high black hole spin. Both
engines should be able to create the radio jets.
Although a viable explanation, from the conceptional view point it
appears somewhat astonishing
that two intrinsically different AGN types should lead to the
{\it same} radio jet and lobe phenomena. 
Alternatively, 
in the picture of orientation-dependent unification,
galaxies could be quasars in which the AGN is hidden 
behind a dust torus seen "edge-on", as proposed by Barthel (1989).
Further debate arises from the fact that galaxies are distributed over
the entire redshift range of the 3CR 
catalogue from the nearby local universe at z\,$\approx$\,0.01 up to z\,$\approx$\,2, 
but the quasars are preferentially found at larger distances at
z\,$\ga$\,0.3 and 178 MHz luminosities above 10$^{\rm 26.5}$ W/Hz
(e.g. Singal 1996, Willott et al. 2000). 
The increase of the quasar fraction with distance and luminosity may find an explanation
by the {\it receding torus model} (Lawrence 1991, Hill et al. 1996):
due to dust sublimation the inner wall of the torus recedes with increasing
luminosity of the central engine 
and -- under the assumption that the torus scale height is independent of
luminosity -- the cone opening angle increases,
leading to a higher chance to see the broad line region.
However, it is still under debate how far this elegant model is valid for the 
powerful radio galaxies with 178 MHz luminosities {\it above} 10$^{\rm 26.5}$ W/Hz.
Alternatively, also physical or cosmic evolution 
could play a role: in this picture galaxies or a subset of them
could be old quasars with a starving black hole, 
but still with luminous radio phenomena, possibly due to a favourable circumgalactic environment. 
Support for an environmental or evolutionary relationship comes
from the fact that
many FR\,2s and most FR\,1s fall into the low ionisation emission line
category (Hine \& Longair 1979), with little evidence for thick obscuring dust tori on HST images
(Chiaberge et al. 1999, 2000).

In order to test the orientation-dependent unification between FR\,2 
  galaxies and quasars, henceforth simply called {\it unification},
the task is to reveal a hidden powerful AGN in a (powerful) 
galaxy. 
Clear evidence for an AGN hidden behind a dust torus 
can be obtained by spectropolarimetry, when broad 
lines show up as light scattered in the bipolar cones 
with the polarisation angle being perpendicular to the polar axis (Antonucci \&
  Miller 1985). 
Despite a number of cases with broad lines revealed by spectropolarimetry, 
in numerous cases huge extinction or dust lanes extended over kpc could
  have prevented the view into the central region via a "suitable
  mirror" (Antonucci \& Barvainis 1990, Hill et al. 1996, Cohen et
  al. 1999).
Note that the
optical signatures of dust extinction only indicate that
something  {\it could} be hidden, but it remains to be shown
that there is actually a powerful energy source behind the dust lanes.
A robust check of the {\it unification} is to look for the mid- and far-infrared
reemission of the absorbed light from the AGN. A great advantage is
that at wavelengths $\lambda$$\ga$25\,$\mu$m the
IR emission is largely optically thin, hence isotropic and
independent of the aspect angle.  

\subsection{Infrared Observations} 
\label{section_status_infrared}

Observations of the 3CR sample with IRAS  
did not allow a substantial
conclusion about the {\it unification}
(Heckman et al. 1992, Heckman et al. 1994, Hes, Barthel \& Hoekstra 1995, 
Hoekstra, Barthel \& Hes 1997).
The wavelength coverage and sensitivity of the IRAS observations were not sufficient to
discriminate between dust and synchrotron emission in the individual sources, 
hence to correct for the contribution by a beamed synchrotron component to
the mid- and far-infrared spectra of the sources. 

With the Infrared Space Observatory ISO  (Kessler et al. 1996) several
small 3CR sub-samples were successfully observed: 
On the basis of a few ISOPHOT-MAMBO detections Haas et al. (1998) demonstrated that 
the IR emission of quasars and galaxies is a mixture of thermal and synchrotron radiation, 
dominated by thermal emission in radio-quiet quasars and synchrotron emission in 
flat spectrum radio-loud ones. 
Remarkably, steep spectrum quasars as well as galaxies showed powerful infrared dust emission. 

In order to test the {\it unification}, a dedicated  
strategy compared galaxy -- quasar pairs, which match in isotropic 
178 MHz radio luminosity as well as redshift (in order to minimize possible cosmological effects).
However, the four FR\,2 galaxy -- quasar pairs 
studied by van Bemmel, Barthel \& de\,Grauuw (2000) yielded a lower
detection rate of thermal dust emission and a lower dust 
luminosity for the galaxies than for the quasars, contrary to the
expectations from the {\it unification}.
On the other hand, ISOPHOT-MAMBO observations of 
ten FR\,2 galaxy -- quasar pairs with high radio power by Meisenheimer
et al. (2001) 
yielded a balanced detection statistics and a similar dust luminosity for the quasars and galaxies,  
as predicted by the {\it unification}. 
Furthermore, the extrapolation of the synchrotron spectra 
of the quasars from the cm and mm regime into the far- and mid-IR has on average 
a more than ten times higher level than
that for the radio galaxies. This is also consistent with the unified schemes, which predict that   
the jet axis of quasars is oriented more toward the line-of-sight, resulting 
in a stronger beamed component. 
Further ISOPHOT-MAMBO studies of small 3CR samples confirmed that 
many 
galaxies as well as quasars produce luminous dust emission (Andreani
et al. 2002), and 
that the synchrotron component in quasars is higher than in
galaxies (van Bemmel \& Bertoldi 2001). 
A closer quantitative look at the dust-to-radio
luminosity ratio, however, reveals also differences between
galaxies and quasars which suggest the additional role of evolution and
environment superposed on the {\it unification}
(Meisenheimer et al. 2001, Haas 2001).

Due to the limited detection of individual 3CR sources by IRAS, no
definite conclusions about the {\it FR\,1-FR\,2 relation} could be drawn;
to our knowledge neither ISO results on that topic have been reported so far. 
\subsection{This paper} 
\label{section_this_paper}

In order to overcome the small number statistics
and derive results from a broader statistical basis, 
we present here the remaining unpublished ISOPHOT
SEDs of the 3CR sources in the ISO archive. 
We discuss the statistical properties of the 
complete mid- to far-infrared data base of 75 galaxies and quasars 
from the 3CR catalogue observed by ISOPHOT with regard to 
unification, evolution and environment as well as the
applicability of the receding torus model.
We focus here on the {\it unification};
the discussion of the {\it FR\,1-FR\,2 relation} will follow in a subsequent paper.
Since common selection criteria for the sources were low cirrus foreground and  
good visibility to the satellite, 
this sample provides a fairly representative sub-sample of the 3CR catalogue, 
with a redshift range from nearby to distant 
(z\,$\la$\,2) objects. 
Throughout this paper we use 
blackbodies modified with a dust emissivity index of $\beta$\,=\,2.
We adopt a $\Lambda$ cosmology with 
H$_0$ = 71 km\,s$^{-1}$\, Mpc$^{-1}$, $\Omega_{{\rm matter}}$ = 0.27
and $\Omega_{\Lambda}$ = 0.73 (Bennett et al. 2003).

\section{The data}
\label{section_data}

The ISO Data Archive (Kessler et al. 2000) contains 5 to 200 $\mu$m
photometry for 75 3CR sources obtained with ISOPHOT  (Lemke et al. 1996), the
photometer on board ISO.
About half of them have also been observed between 5 and 15 $\mu$m 
with ISOCAM. 
Our sample does not contain those 3CR sources, for which only 
ISOCAM mid-infrared data are available, without any far-infrared complement
by ISOPHOT data or IRAS detections. The full set of 85 ISOCAM
observations is published by Siebenmorgen et al. (2004). 
Our sample does not include the two FR\,1 sources 3C\,71 (NGC\,1068) and
3C\,231 (M\,82), which do not belong to the classical FR\,1s with
elliptical hosts, and which are nearby and therefore
too extended to be observed in ISOPHOT photometry modes. 
In order to characterize the (sub)-mm part of the SEDs 
for the full sample, we complemented the available literature data: 
For 15 sources we found submm photometry in the JCMT-SCUBA archive, 
and performed additional mm observations for 28 sources with the 
Max-Planck Millimeter bolometer array MAMBO (Kreysa et al. 1998) at the IRAM 30\,m telescope.  

\subsection{ISOPHOT}
\label{section_isophot}

The observing modes (Laureijs et al. 2002) 
comprise chopped measurements, including those with the spectrometer ISOPHOT-S, 
from which we determined broad band fluxes, and small maps. 

\begin{figure*}
\vspace{-1.9cm}
\begin{center}
\hspace{0.5cm}
   \epsfig{file=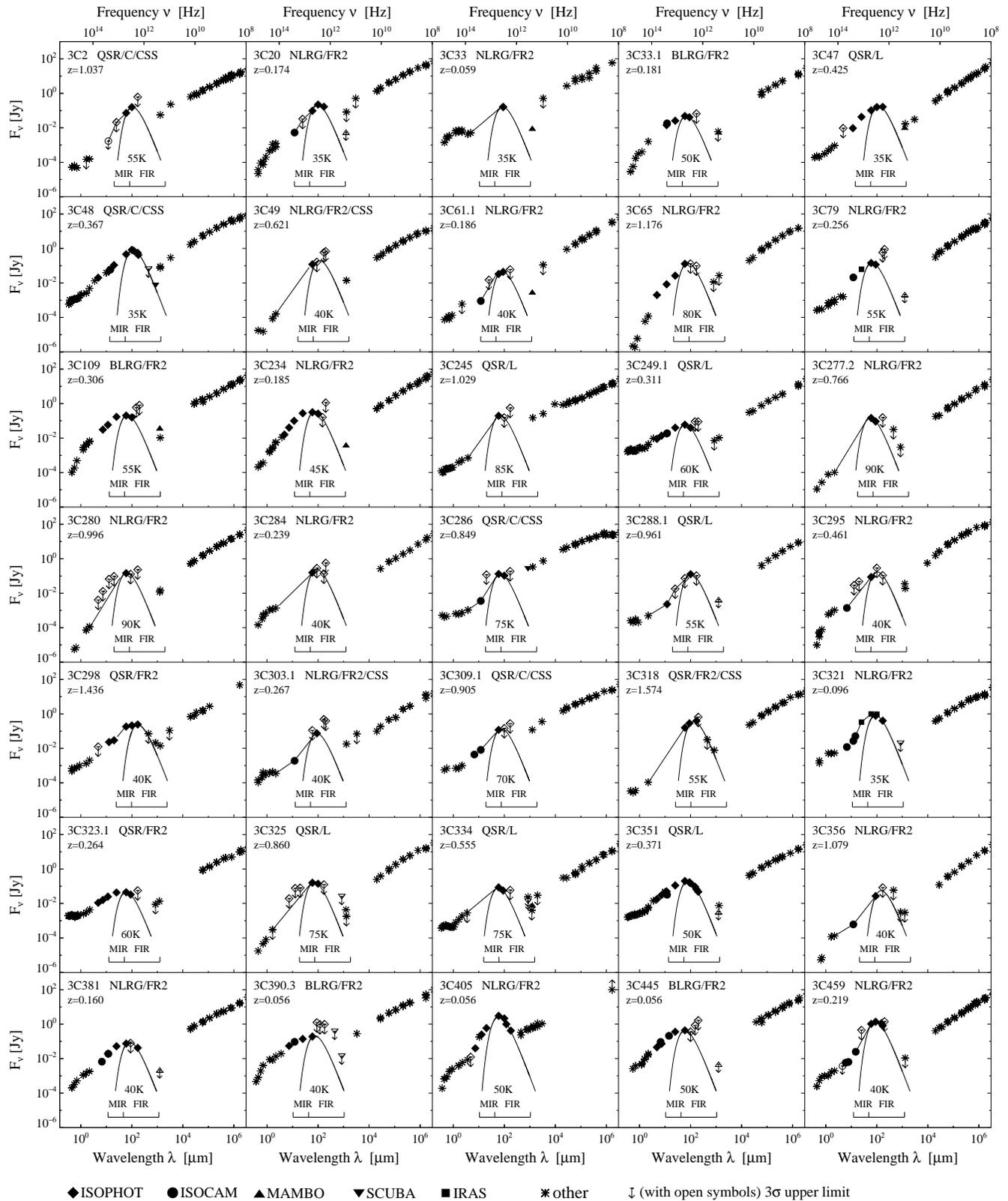,width=16.5cm}
\end{center}
\vspace{-1.5cm}
\caption[]{\label{msxxxx_fig_seds} 
  Spectral energy distributions of those quasars, BLRGs and FR\,2
  NLRGs which show a clear thermal bump and are used in the
  discussion of the unification in
  Sect.\,\ref{section_statistical_properties}.
  The measurement errors are of the size of the symbols.
  The wavelength and frequency ranges are as observed and not 
  corrected to the rest frame of the objects.
  The MIR (10-40$\mu$m) and FIR (40-1000$\mu$m) wavelength ranges, as
  well as the temperature of the modified blackbody (solid line curve) fitted to the
  thermal FIR bump are shown in the object's restframe. Where the data coverage is
  sparse, the straight lines indicate the "envelopes"
  used for computing the restframe luminosities.
}
\end{figure*}

\begin{figure*}
\begin{center}
  \vspace{-0.3cm}
  \hspace{-1.2cm}
  \epsfig{file=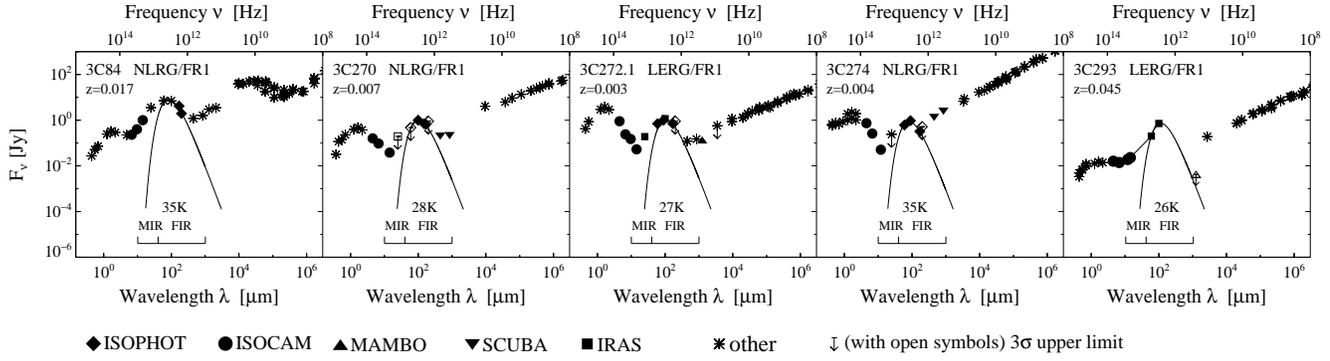,width=19cm,clip=true}
\end{center}
\vspace{-0.9cm}
\caption[]{\label{msxxxx_fig_seds_2} 
  Spectral energy distributions of the FR\,1 sources.
  The meaning of the lines and labels is as in Fig.\,\ref
  {msxxxx_fig_seds}.
  Where the ISOPHOT data coverage is sparse, 
  IRAS flux values from the NED are shown, if available (e.g. for 3C\,293). 
  }
\end{figure*}

\begin{figure*}[t!]
\vspace{-1.9cm}
\begin{center}
  \hspace{0.5cm}
  \epsfig{file=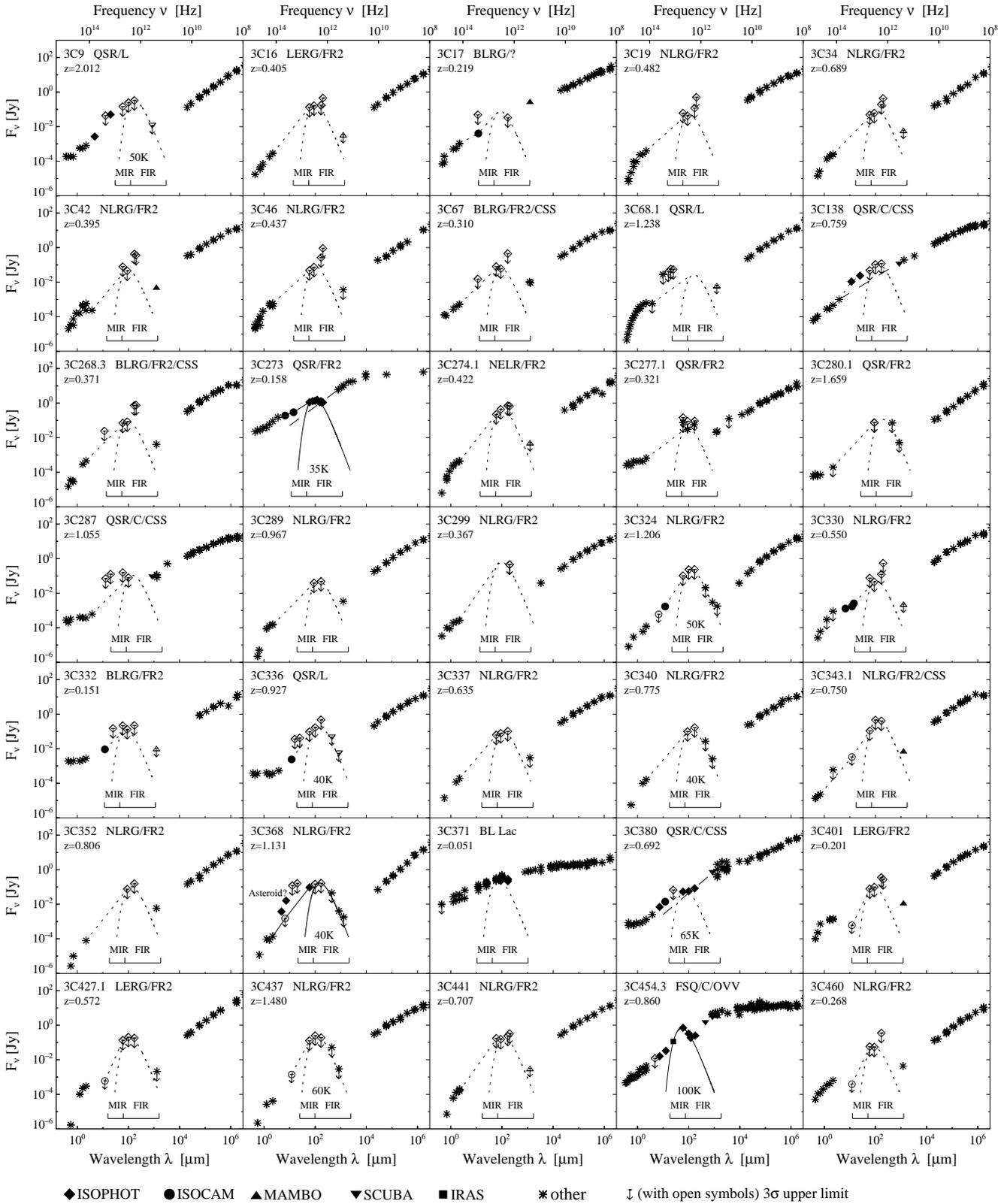,width=16.5cm}
\end{center}
\vspace{-1.5cm}
\caption[]{\label{msxxxx_fig_seds_3}
  Spectral energy distributions of the remaining sources not used in
  the discussion, which are
  either flat spectrum sources or have 
  only poorly constrained thermal bumps.
  The meaning of the lines and labels is as in Fig.\,\ref
  {msxxxx_fig_seds}. For the poorly constrained cases, however, the blackbody curves and the envelopes
  are shown here as {\it dotted} lines in order to emphasize that they
  represent only a maximum possible dust contribution. For the
  blackbody, if not sufficiently constrained
  by the data, we adopted the average of T$_{\rm p}$ $\approx$ 40 K derived
  from the good cases shown in Fig.\,\ref{msxxxx_fig_seds}. 
  The long dashed lines for 3C\,138, 3C\,273 and 3C\,380 indicate the extrapolation of the
  synchrotron emission towards shorter wavelengths.
  }
\end{figure*}

We have reduced the data using the ISOPHOT Interactive Analysis tool
(PIA\footnote{PIA is a joint development by the ESA Astrophysics 
Division and the ISOPHOT consortium led by the MPI f\"ur Astronomie,
Heidelberg. Contributing institutes are DIAS, RAL, AIP, MPIK, and MPIA.}
V10), together with the calibration data set V7.0. 
Using the latest versions of the data reduction tools, 
there are no systematic photometric offsets with regard to PIA V7 and V8 used 
for most of the results published earlier by various authors. The accuracy
of the absolute photometric calibration depends mainly on systematic errors
as described in Laureijs $\&$ Klaas (1999), and it is
currently known to be better than 30\% for faint sources. 
The relative filter-to-filter calibration agrees within 15\% 
(Klaas et al. 2002). 

\subsection{MAMBO}
\label{section_mambo}

With the MAMBO array at the IRAM 30-m telescope on Pico Veleta, Spain, during the
pooled observation campaigns between January and March 2003, we obtained
1.2\,mm (250 GHz) continuum flux densities for 28 of our sources.  We
used the standard on-off photometry observing mode, chopping between
the target and sky at 2 Hz, and nodding the telescope every 10 s.
The total on sky integration time varied between 2.5 and 23
minutes per source, and each object was observed at least twice, on
different days.
The absolute flux calibration was established by observations of
Mars and Uranus, resulting in a flux calibration uncertainty of
approx. 20\%.  The data were reduced using the MOPSI software package.

\subsection{SCUBA}
\label{section_scuba}

Yet unpublished submm continuum data obtained with SCUBA were
retrieved from the JCMT archive for 15 sources. The observations were
performed in photometry mode, and for one source (3C\,274 = M\,87) as
jiggle maps. We have reduced them with the SCUBA User Reduction
Facility (SURF), including identification of noisy bolometer pixels
and removal of sky noise.  The atmospheric transmission was determined
from skydips and water radiometer data at the CSO. IRC+10206 and
HL-Tau served as standard calibrators. The integration time per source
was 10-20 min.  In case of very good atmospheric transmission
($\tau$$_{\rm CSO}$\,$\approx$\,0.05) we also included the 450 $\mu$m
data.  The photometric accuracy at 450 $\mu$m and 850 $\mu$m is about
30\% and 25\%, respectively.

\subsection{Literature data}
\label{section_literature_data}

ISO photometry of 49 3CR sources has so far
been published by various authors, as labelled in col. ``publ'' of Table\,\ref{msxxxx_tab_fluxes}.
The SED plots shown here include 4.5 to 14.3 $\mu$m data with ISOCAM 
(Cesarsky et al. 1996) for 33 sources, which are published 
by Siebenmorgen et al. (2004) and for a subset by Freudling et al. (2003). 

Additional literature data were retrieved from the NED and SIMBAD, 
the 2MASS archive and numerous papers:
  Lilly \& Longair (1984),
  Spinrad et al. (1985), Laing et al. (1983), Lilly et al. (1985), 
  Neugebauer et al. (1987), Simpson et al. (2000)
  and de\,Vaucouleurs et al. (1991) at optical and NIR wavelengths, 
  Chini et al. (1989a,b),
  Steppe et al. (1988, 1992, 1993),
  Robson et al. (1998),
  Hughes et al. (1993),
  Best et al. (1998), van Bemmel et al. (2000),
  Polletta et al. (2000),
  Archibald et al. (2001),
  Willott et al. (2002),
  Andreani et al. (2002) and
  Stevens et al. (1998) 
  at mm wavelengths, and at radio wavelengths from Jodrell-Bank Web
  pages (at http://www.jb.man.ac.uk/atlas/),
  K\"uhr et al. (1981), 
  Kellermann et al. (1989) and
  Akujor et al. (1994). 

\section{Results}
\label{section_results}

\subsection{General properties}
\label{section_general_properties}

The IR, submm and mm fluxes of the 3CR sources are listed in Table\,\ref{msxxxx_tab_fluxes}. 
ISOPHOT achieved a detection rate of about 70\%
in the MIR ($\le$\,25\,$\mu$m) and  60\% in the FIR ($\ge$\,60\,$\mu$m). 
Even at the long wavelengths between 120 and 200\,$\mu$m the detection 
rate is about 20\%.
The fluxes of the brightest non-variable sources, also detected by IRAS 
(e.g. 3C\,272.1, 3C\,405, 3C\,459), agree within 15\%. 

14 out of 28 sources observed with MAMBO at 1.2\,mm  were
detected, and 3-$\sigma$ upper flux limits are provided for the others. 
From the 15 sources in the SCUBA archive eight were detected
at 850\,$\mu$m  and three  at 450\,$\mu$m. 

Supplementing our data with further photometry from the 
literature as listed in section\,\ref{section_literature_data},
we display the observed spectral energy distributions (SEDs)
in Figs.\,\ref{msxxxx_fig_seds}, \ref{msxxxx_fig_seds_2} and \ref{msxxxx_fig_seds_3}, whereby 
each Figure contains a subset of sources suitably chosen with regard
to the illustration of the discussion items.
Fig.\,\ref{msxxxx_fig_seds} shows only those quasars, BLRGs and FR\,2
NLRGs which exhibit a clear thermal bump and are used for 
investigating the {\it unification}.
Fig.\,\ref{msxxxx_fig_seds_2} shows the FR\,1 SEDs. Finally, 
Fig.\,\ref{msxxxx_fig_seds_3} shows the remaining SEDs which are
  either those of flat spectrum sources or which exhibit  
  poorly constrained thermal bumps due to upper limits only --
  nevertheless illustrating the high IR luminosity limits.
Note that the SEDs in Figs.\,\ref{msxxxx_fig_seds}
and \ref{msxxxx_fig_seds_2}  run smoothly and the photometric uncertainties
represented by the size of the symbols
are small.

The most remarkable feature of the SEDs in Fig.\,\ref{msxxxx_fig_seds}
is the steady increase in flux density 
from near- to far-IR wavelengths with a peak typically at 60-100\,$\mu$m,  
followed by a steep drop at (sub)-mm wavelengths in many cases. Where
it could be measured, in
particular for 3C\,48, 3C\,298, 3C\,318, 3C\,405, this
drop follows exactly the Rayleigh-Jeans tail of the modified blackbody
(with dust emissivity index $\beta$\,=\,2). This strongly 
suggests a thermal nature of the FIR emission.

In order to discuss the SEDs in the framework of thermal emission, we
have eyeball fitted a modified 
blackbody to the {\it peak} of
the flux distribution, as shown in Figs.\,\ref{msxxxx_fig_seds}
and \ref{msxxxx_fig_seds_2}. 
The peak blackbody is mainly determined by the maximum in the FIR
and, if available, also by the sub-millimetre points.
However, this greybody fit does not always match the
(sub)-millimetre measurements, since they may often be dominated
by the synchrotron emission (for example in 3C\,109 or 3C\,234 in
Fig.\,\ref{msxxxx_fig_seds}).
Note that our "peak blackbody" corresponds in a $\nu$F$_\nu$ versus $\nu$ plot
to the "break" of the thermal spectrum, where the slope equals unity.
Its temperature T$_{\rm p}$ (corrected for redshift) lies between  $\sim$30 and
100\,K, with higher T$_{\rm p}$ values preferentially found in
the most luminous sources at high redshift - probably a natural 
consequence for a flux limited sample like ours.
It is clear from Fig.\,\ref{msxxxx_fig_seds} that also dust which is
colder than the peak blackbody
may be present, but if so, then at lower flux levels. Also, as being evident
from the MIR data points above the peak blackbody curve, warmer dust
components up to the dust sublimation temperature of about 1500 K
are present, but for lucidity of the SED plots they are not shown explicitly here. 
Note that for the cases of upper limits and poorly determined dust
bumps shown in Fig.\,\ref{msxxxx_fig_seds_3} we tentatively
adopt an average T$_{\rm p}$ $\sim$ 40 K determined from the good
cases shown in Fig.\,\ref{msxxxx_fig_seds}, if not constrained sufficiently.

The restframe thermal mid- (10-40 $\mu$m)  and far-IR (40-1000 $\mu$m) luminosities
are computed by integrating the SED "envelopes". These curves are either
sufficiently sampled by  observations, as e.g. in the case of 3C\,48, or they are
partly sampled between 10 and 100 $\mu$m and follow the
Rayleigh-Jeans tail of the peak blackbody between 100 and 1000
$\mu$m, as e.g. in the case of 3C\,47.
For a few sources in Fig.\,\ref{msxxxx_fig_seds} (3C\,33, 3C\,49, 3C\,245 and 3C\,284) the SEDs
are sparsely sampled in the MIR-FIR regime; in these cases 
they are interpolated linearly in the log-log SED plots between NIR and FIR
flux points as indicated by the solid lines. 
Since the short wavelength part of the integration interval dominates the
luminosity calculation, for poorly sampled SEDs the MIR luminosity may be
overestimated by up to $\sim$50\%, but the FIR luminosity is not significantly ($<$15\%)
affected by the choice of T$_{\rm p}$. 

For all sources Table \ref{msxxxx_tab_luminosities} lists the derived
parameters like luminosities, dust temperatures and masses, 
and other values used in the discussion.

\subsection{Notes on some individual sources}
\label{section_notes}

{\it 3C\,274 (M\,87): }
The SED of this FR\,1 source clearly exhibits as three basic features
(1) the optical-NIR bump from
the host galaxy, (2) the thermal dust bump in the mid- and far-IR, and
(3) the
radio synchrotron component. 
Table\,\ref{msxxxx_tab_fluxes} lists the total fluxes at 450
and 850$\mu$m, nevertheless we were able to resolve
the one-sided jet and the core in the SCUBA jiggle maps. The core
fluxes assuming a point source with FWHM 8$\arcsec$ and 15$\arcsec$
are 0.391$\pm$0.10 Jy and 1.085$\pm$0.10 Jy at 450 and 850 $\mu$m, respectively.  

{\it 3C\,268.4:} 
Our FIR fluxes at 90 and 170 $\mu$m are similar to those reported by Andreani et al. (2002), 
but they are surprisingly high for a quasar at z\,$\approx$\,1.4. 
Therefore we checked for the presence of other nearby galaxies. 
In fact, the galaxy NGC\,4138 lies less than 200$\arcsec$ apart and has an extent of more than 
100$\arcsec$. In the FIR it is also bright with F(170$\mu$m) $>$ 6 Jy, as we found 
with the help of the ISOPHOT Serendipity Survey 
(Stickel et al. 2000). 
Hence 
the large ISOPHOT beams of 45$\arcsec$ and 180$\arcsec$, respectively, 
have seen mainly the contamination by NGC\,4138.
Therefore, we do not consider 3C\,268.4 any further.

{\it 3C\,313:} The FIR flux reported by Andreani et al. (2002) is due
to a misidentification, 
actually the ISO observations refer to IRAS F15086+0801 ($\ne$ 3C\,313).

{\it 3C\,368:} The source was observed three times in the  MIR 5-15\,$\mu$m range,  
once with ISOPHOT on Oct. 18, 1996 (at 4.8 and 7.3\,$\mu$m) and twice  at 12\,$\mu$m with 
ISOCAM on Feb. 27, 1996, and on Oct. 27, 1997. 
The detections on Oct. 18, 1996 exceed by far the 3$\sigma$ upper limits 
from Feb. 27, 1996 and Oct. 27, 1997. 
The ISOPHOT chopped series show a perfectly oscillating  on-off pattern leaving no doubt 
about the reality of the source detection. 
Since the explanation of strong 
MIR variability of 3C\,368 seems unlikely to us, we suggest that the ``MIR flux excess`` 
during the observations on Oct. 18, 1996 
is due to the passage of an asteroid (or comet), seen in the ISOPHOT aperture. 
However, no so far known asteroid has crossed the ISOPHOT observations 
(private communication by Thomas M\"uller).
Nevertheless, an asteroid (or comet) 
seems to be the best explanation for the spurious 3C\,368 detection and non-detections.
Furthermore, at a temperature of 230\,K as indicated by the MIR
fluxes,
an asteroid superposed on the 3C\,368
field of view does not affect 
the FIR fluxes of 3C\,368 at $\lambda$ $\ge$ 60\,$\mu$m. 
Since 3C\,368 was also detected with SCUBA at 450\,$\mu$m (Archibald et al. 2001), 
we ascribe the FIR-submm dust bump to 3C\,368 and use the ISOCAM upper limit for 
MIR luminosity estimates. Nevertheless, we do not include 3C\,368
in the discussion here.

{\it 3C\,390.3:} This broad line radio galaxy was also detected by IRAS, 
but both the ISOPHOT and the ISOCAM MIR data do not confirm the exceptionally 
high 12 and 25 $\mu$m IRAS fluxes (Miley et al. 1984), thus making the mid- to 
far-IR colours of 3C\,390.3 more similar to those of other BLRGs. The
high IRAS fluxes could be contaminated by an asteroid or comet.

{\it 3C\,454.3:} 
This is one of the few examples of a {\it flat} spectrum quasar with a
strong FIR dust bump
sticking out of the bright synchrotron continuum. Other such examples
are 3C\,273 (Fig.\,\ref{msxxxx_fig_seds_3}), 3C\,279 (Haas et al. 1998)
and PG\,1302-102 (Haas et al. 2000).
We checked, whether the bump in 3C\,454.3 could be due to another galaxy, 
but the only candidate on DSS and 2MASS images is an unresolved object 
about 15$\arcsec$ NNE with a K-star like optical-NIR SED. 
Therefore we conclude that the bright FIR dust bump is in fact
caused by 3C\,454.3.
\section{Discussion}
\label{section_discussion}

Figs.\,\ref{msxxxx_fig_seds} and \ref{msxxxx_fig_seds_2} demonstrate
that at least half ($>$ 40/75) of the
3CR SEDs display a mid- to far-infrared bump due to thermal emission by dust. 
Even in the remaining cases the upper limits allow for considerable
dust emission 
(Fig.\,\ref{msxxxx_fig_seds_3}, Table\,\ref{msxxxx_tab_luminosities}).
In the following we consider the different classes (e.g. flat and steep spectrum quasars, BLRGs, 
FR\,2 NLRGs and FR\,1s), and investigate
how the dust emission compares to the properties at other
wavelengths. 
First, we discuss the SED shapes for representative class members.
Then we study the statistical properties of the FR\,2 NLRGs,
quasars and BLRGs with respect to the 
{\it unification}, whereby we also
consider other effects like those predicted by
the receding torus model as well as  evolution and environment.
\subsection{SED shapes}
\label{section_basic_sed_shapes}

The SEDs shown in Figs.\,\ref{msxxxx_fig_seds} and \ref{msxxxx_fig_seds_2}
exhibit a striking diversity of shapes. 
Figure\,\ref{msxxxx_fig_sed_shapes} shows five representative examples
among the different AGN classes, i.e. radio and optical types:

\begin{figure}
  \vspace{0.95cm}
  \hspace{1.5cm}
      \epsfig{file=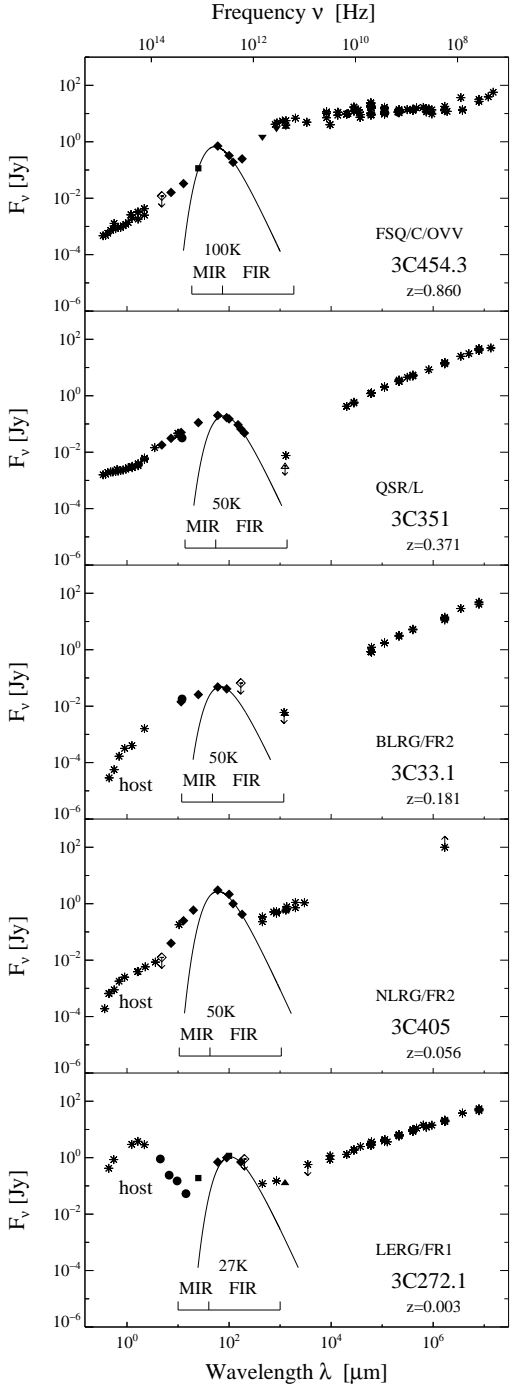, width=11.25cm}
\vspace{0.95cm}
\caption[]{
\label{msxxxx_fig_sed_shapes} 
SED shapes of 3CR sources as representative examples of the different optical
  and radio classes, having good spectral data coverage. The symbols
  are as in Fig.\,\ref{msxxxx_fig_seds}.
  }
\end{figure}

1) Flat spectrum core dominated quasars 
(``FSQs'') like 3C\,454.3: 
The synchrotron spectrum extends at a nearly constant flux level 
from the cm to the mm range, and declines at shorter wavelengths. 
In the FIR, the SEDs may show a thermal dust bump, as in the powerful FSQ 
3C\,454.3, or not, as in the less luminous BL Lac object 3C\,371
(Fig.\,\ref{msxxxx_fig_seds_3}). 
In the context of unified schemes these flat spectrum 
sources are viewed nearly pole-on as proposed by Orr \& Browne (1982).

2) Steep spectrum quasars like 3C\,351: 
The synchrotron spectrum drops from the cm to the mm wavelengths, 
so that the thermal FIR dust emission bump becomes clearly visible.
The optical spectra are quite blue, 
typical for an AGN outshining the host galaxy. 
Also, the SEDs show a power-law like flux increase, from the NIR 
over the MIR and peaking in the FIR, as seen also in  optically identified PG 
quasars (Haas et al. 2003).
With respect to the optical flux, the FIR dust bump may be quite
pronounced with a steep slope from the NIR to the FIR as in 3C\,48 at
z$\sim$0.3 and
3C\,318 at z$\sim$1.5 or more shallow as in 3C\,249.1 at
z$\sim$0.3 and 3C\,298 at z$\sim$1.5 (Fig.\,\ref{msxxxx_fig_seds}).
We do not see any difference between the SEDs of FR\,2, lobe dominated
and CSS types.

3) Broad line radio galaxies (BLRG) like 3C\,33.1: 
The {\it infrared} and {\it radio} spectra are similar
to those of the quasars, 
showing in the IR also the powerlaw rise between 5 and 60 $\mu$m. 
In contrast, the {\it optical} spectrum is redder than that of quasars, probably
due to
a stronger contribution of the host galaxy relative to that of the AGN,
and possible obscuration of the intrinsically blue central spectrum. 

4) Narrow line radio galaxies (NLRG) of class FR\,2 like 3C\,405:
The far-infrared and radio spectra are similar to those of quasars and BLRGs.
Relative to the FIR, however, the emission in the optical and also in the NIR is  
weaker. This suppression at short wavelengths shows some range:  
while for 3C\,405 the optical-NIR SEDs are only ``moderately'' reddened, 
the optical to IR slope is very steep for both 3C\,65 and 3C\,234 (Fig.\,\ref{msxxxx_fig_seds}). 
In the unified scheme the NLRGs of class FR\,2 represent edge-on quasars.
3C\,234 shows polarised broad H$_{\alpha}$ but narrow H$_{\beta}$ lines 
(Antonucci 1984); hence it might be seen at the grazing angle to the edge of
the dust torus.
Some NLRGs of class FR\,2
like  3C\,321 and 3C\,459 (Fig.\,\ref{msxxxx_fig_seds}) show
a strong FIR bump and only a moderate MIR contribution, 
more reminiscent of the SEDs of starburst galaxies. 

5) Narrow line and low-excitation radio galaxies (LERG) of
class FR\,1 like 3C\,272.1,  
showing three basic SED features: 
\begin{itemize}
\item[1)] in the optical-NIR a bright bump due to the elliptical hosts,
  represented by a blackbody of 
T\,$\approx$\,3500\,K peaking at about 1\,$\mu$m, 
\item[2)] in the cm-mm wavelength range a steep synchrotron spectrum, and  
\item[3)] in the MIR-FIR a well discernible dust emission bump.
\end{itemize}

The examples shown in Fig.\,\ref{msxxxx_fig_sed_shapes}
illustrate the basic SED features. In the optical
they arise from the host galaxy and the blue AGN continuum, in some
cases reddened by extinction toward the central region.  In the IR and radio
ranges the emission from dust and from the jet and lobes can well be
discerned. Remarkably,
the SEDs exhibit also a diversity of shapes even within each AGN class, 
for example in the strength of the FIR dust bump or the slope between
the NIR and FIR fluxes. For the optically selected PG quasars, which
are thought to be seen nearly pole-on, such differences in the
IR emission could be interpreted as
evolution of the dust distribution around the
heating sources (Haas et al. 2003).
Although the diversity of SED shapes within each 3CR AGN class may
also indicate such intrinsic differences, 
the orientation with respect to the line of sight is not yet
known for our 3CR sample and it is still under debate, how far the observed near- and mid-IR
emission is influenced by different viewing angles.
Therefore, we will restrict our investigation to the orientation problem.
Furthermore, average SEDs for each AGN class could
be computed, but they will smear out the details recognisable in the
individual SEDs.
Therefore, instead of using averages, in the next subsection we prefer to analyse the
properties from the individual SEDs. 

\subsection{On the FR\,2 unification}
\label{section_statistical_properties}

In this section we consider the following two basic classes:
i) the steep spectrum  quasars and the BLRGs, henceforth for short denoted as {\it
  quasars}, and ii) the  FR\,2 NLRGs, henceforth denoted as {\it
  galaxies}. The strategy to check the {\it unification} includes two steps:
(1) to show that both the {\it quasars} and the {\it galaxies} exhibit
a high mid- to far-IR luminosity ratio
typical for AGNs, and (2) that the isotropic
FIR-to-radio luminosity ratio is the same for {\it quasars} and {\it galaxies} at
matched isotropic 178 MHz radio power.
We consider only those
sources
shown in Fig.\,\ref{msxxxx_fig_seds}, which have 
sufficiently well sampled SEDs.
The sample investigated here consists of 17 {\it galaxies} and 18
{\it quasars}, hence it can be considered as balanced.
The set of sources shown in Fig.\,\ref{msxxxx_fig_seds_3} and 
not included here comprises  20 {\it galaxies}  and 15 - partly flat
spectrum - {\it
  quasars}, 
and we will see below that their exclusion does not introduce a strong bias.

\subsubsection{AGN typical dust emission}
\label{section_agn_typical_dust_emission}

\begin{figure}
  \hspace{-0.7cm}
   \epsfig{file=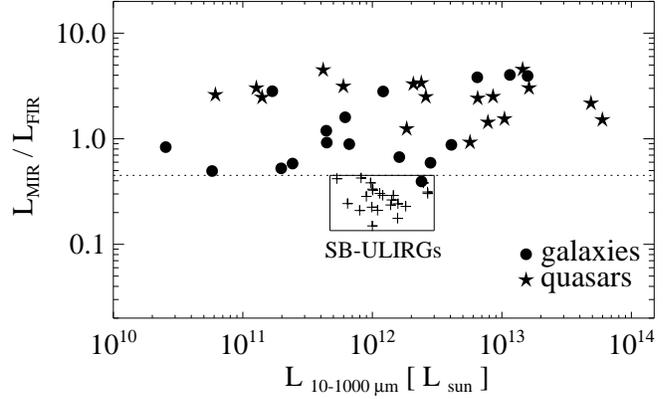 ,width=9.8cm, clip=true} 
\caption[]{ \label{msxxxx_fig_lmir_to_lfir_vs_lir} 
Mid- to far-infrared dust luminosity ratio versus 10-1000$\,\mu$m luminosity. 
The rectangular box shows the range of local starburst ULIRGs (from Klaas et
al. 2001), each plotted with a "+" sign,  
and the dotted line indicates the upper end of their L$_{\rm MIR}$ /
L$_{\rm FIR}$ distribution. The only 3CR source falling below this
line is 3C\,459.
}
\end{figure}

The thermal 10-1000 $\mu$m infrared
luminosities of our sample span four orders of  
magnitude, from moderate 10$^{\rm 10}$\,L$_{\odot}$ over luminous 
10$^{\rm 11}$\,L$_{\odot}$  and ultraluminous 10$^{\rm 12}$\,L$_{\odot}$ to 
hyperluminous 10$^{\rm 13}$-10$^{\rm 14}$\,L$_{\odot}$ objects 
(Table\,\ref{msxxxx_tab_luminosities} and abscissa of Fig.\,\ref{msxxxx_fig_lmir_to_lfir_vs_lir}). 
The most luminous objects lie at the highest redshift, 
as is typical for flux limited samples like the 3CR catalogue.

The dust masses derived from the peak blackbodies
shown in Fig.\,\ref{msxxxx_fig_seds} 
range between 10$^{\rm 6}$ and 10$^{\rm 9}$\,M$_{\odot}$ 
(Table\,\ref{msxxxx_tab_luminosities}), which is  
comparable to those found in PG
quasars (Haas et al. 2003) and in ULIRGs (Klaas et al. 2001). 
Although the derived dust masses are possibly uncertain by a factor of
ten, they do still provide a rough estimate of
interstellar material associated with these objects. 

In order to compare the 3CR sources with other samples of known properties,
we have decomposed the IR emission into the 
MIR (10-40$\,\mu$m) and FIR (40-1000$\,\mu$m) 
luminosities (Table\,\ref{msxxxx_tab_luminosities}).
For the {\it quasars} as well as for the {\it galaxies} the mid- to far-infrared
luminosity ratio ranges between 0.4 and 5 as shown in 
Fig.\,\ref{msxxxx_fig_lmir_to_lfir_vs_lir}. 
This is comparable to what is found for optically selected PG quasars (Haas et al. 2003).
In contrast, those ultraluminous infrared galaxies (ULIRGs), in
which the dust emission is powered  mainly by starbursts and {\it not} by an AGN,
populate the L$_{\rm MIR}$/L$_{\rm FIR}$ range from 0.15 to 0.5 (Klaas et al. 2001).
This is significantly lower than that of both the {\it quasars} and  the {\it galaxies}.
Therefore, the high mid- to far-infrared 
luminosity ratio above 0.5  
together with
the high dust luminosity (L$_{\rm IR}$ $\ga$ 10$^{\rm
  11}$\,L$_{\odot}$)
provides evidence for the presence of a powerful AGN in the
{\it quasars} and the {\it galaxies} as well.  
Their AGN may be accompanied by starbursts, but for our purpose here
the presence of a powerful buried AGN in the {\it galaxies} is relevant.
It argues
in favour of the {\it unification}.
It should be noted here, that the orientation-dependent unification picture may not apply for the FR\,1 
and the low-power FR\,2 sources,
which have relatively low dust luminosities and masses. 

A closer look on the mid-to-far-infrared luminosity ratio shows that 
the {\it quasars} lie 
above unity, while many of the {\it galaxies} lie below unity
(Fig.\,\ref{msxxxx_fig_lmir_to_lfir_vs_lir}). 
Such a difference between {\it quasars} and {\it galaxies} is consistent with unified models, 
if the proposed dust torus becomes optically thick at MIR wavelengths
(Pier \& Krolik 1992, 1993, van Bemmel \& Dullemond 2003). In this case, if seen edge-on,
our view towards the hot MIR emitting dust at the inner walls of the torus will
suffer from extinction even at MIR wavelengths by the outer parts of
the torus or by more extended dust. 
The actual degree of MIR opacity is not yet known.
A value of A$_{MIR}$ $\sim$ 1-2 is needed to explain
the lower L$_{\rm MIR}$/L$_{\rm FIR}$ ratios for the {\it galaxies}. Using standard
extinction curves this value corresponds to A$_{V}$ $\sim$
100-200, hence not exceptional with regard to that found in ULIRGs
(e.g. Haas et al. 2001). 
Therefore we do not yet see any conflict with the {\it unification} and
further investigations with more detailed observations
as well as models may clarify that issue.

\subsubsection{Radio--Infrared--Comparison}
\label{section_radio_infrared}

\begin{figure}
  \hspace{-0.7cm}
    \epsfig{file=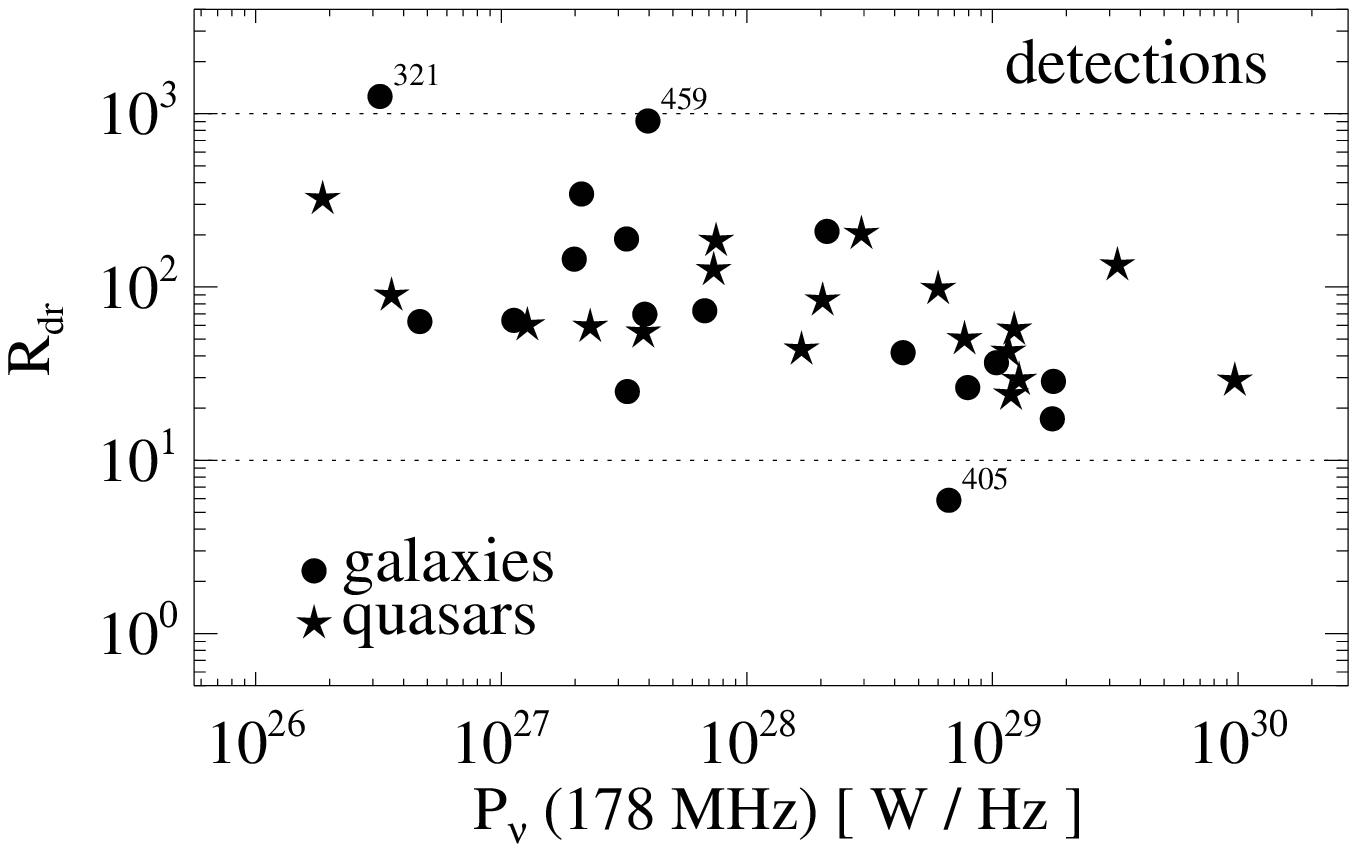  ,width=9.8cm, clip=true}
    
  \hspace{-0.7cm}
    \epsfig{file=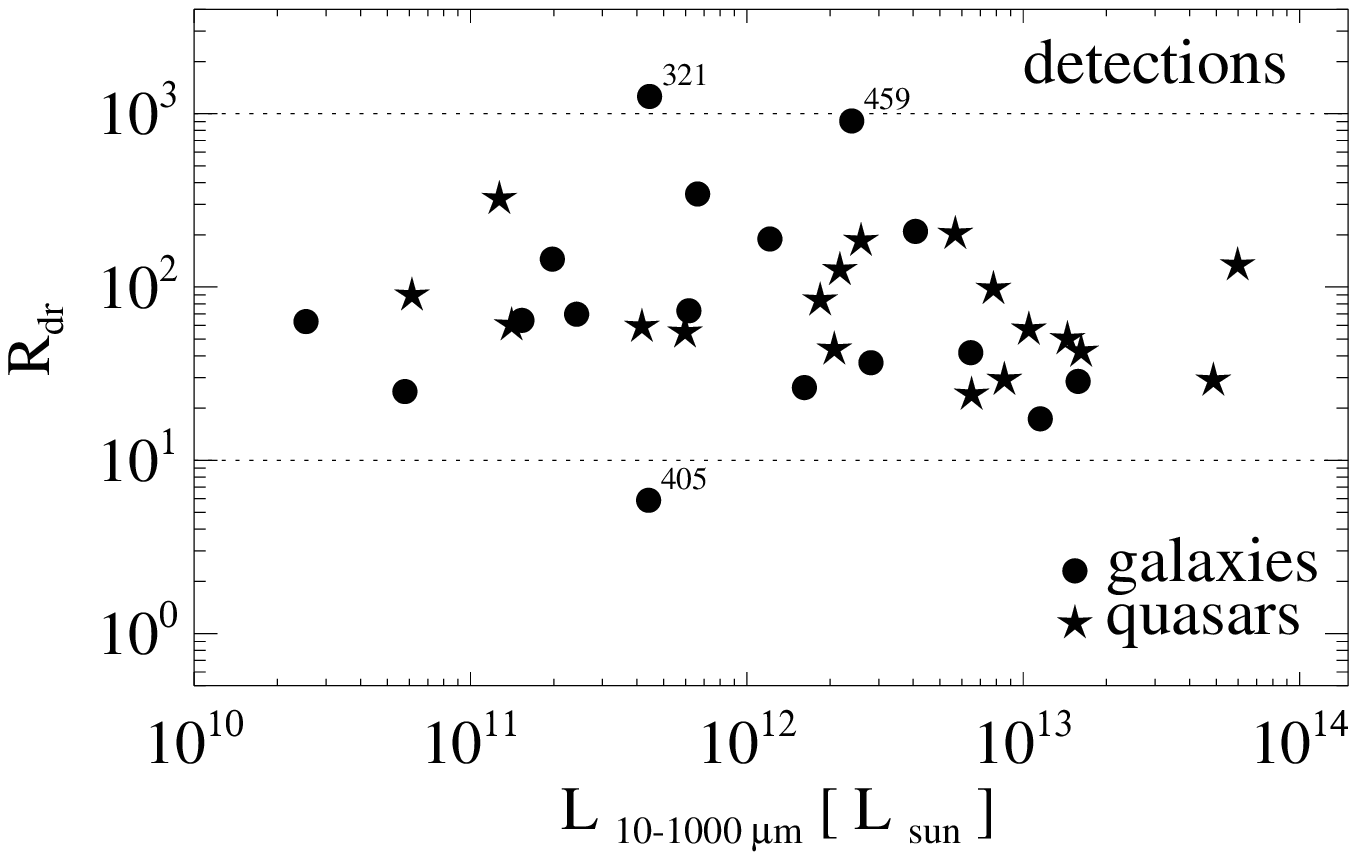  ,width=9.8cm, clip=true}
  \caption[]{ \label{msxxxx_fig_p80_to_p178_vs_z} 
    {\it Top:} Ratio of far-infrared dust power to radio lobe power {\it R$_{dr}$}
    versus 
    radio lobe power,  at restframes
    70 $\mu$m and 178 MHz.
    {\it Bottom:} {\it R$_{dr}$} versus {\it L$_{IR}$}.
    The dotted lines indicate the range of the dispersion discussed in
    the text, and the extreme sources 3C\,321, 3C\,405 and 3C\,459 are
    labelled explicitly.
  }
\end{figure}

So far we have found evidence for a powerful AGN in both the {\it quasars}
and the {\it galaxies}. In a strict sense, however,
the concept of {\it unification} requires that for an object drawn
from the parent population any 
isotropic emission remains the same while rotating the viewing angle to
its axis.
Thus, for an ideal sample of parent objects 
the emitted isotropic FIR dust power should be the same
for objects of identical isotropic lobe power. 
Therefore, following the ideas by Meisenheimer et al. (2001),
we consider R$_{\rm dr}$, the ratio of  P$_{\rm \nu}$
at FIR wavelength 70 $\mu$m (= 4.3 THz) 
and P$_{\rm \nu}$ at radio frequency 178 MHz.
Figure\,\ref{msxxxx_fig_p80_to_p178_vs_z} shows
R$_{\rm dr}$  versus the 178 MHz radio lobe power.
All along the range of the 178 MHz radio lobe power, the distribution of
R$_{\rm dr}$ for the {\it quasars} is strikingly similar to that of the
{\it galaxies}.
There is a marginal impression  of a decline of R$_{\rm dr}$ with 178 MHz radio lobe power. 
This trend, however, is most likely caused by the mathematical dependency "a/x versus x".
It disappears when plotting R$_{\rm dr}$
versus L$_{\rm IR}$ (Fig.\,\ref{msxxxx_fig_p80_to_p178_vs_z} bottom).
As a quantitative check, we divide the luminosity range into two bins,
one for luminosity below 2$\cdot$10$^{\rm 12}$ L$_{\odot}$ and one
above, and calculate for each bin the mean R$_{\rm dr}$ values.
They lie in the range of 10$^{\rm 2.0 \pm 0.3}$ (Table\,\ref{msxxxx_table_rdr}). 
Again, a possible marginal decline of the mean R$_{\rm dr}$ with increasing luminosity is
not yet statistically significant
and we discuss possible reasons further below (Sect.\,\ref{section_other_influences}).
\begin{figure}
  \hspace{-0.7cm}
    \epsfig{file=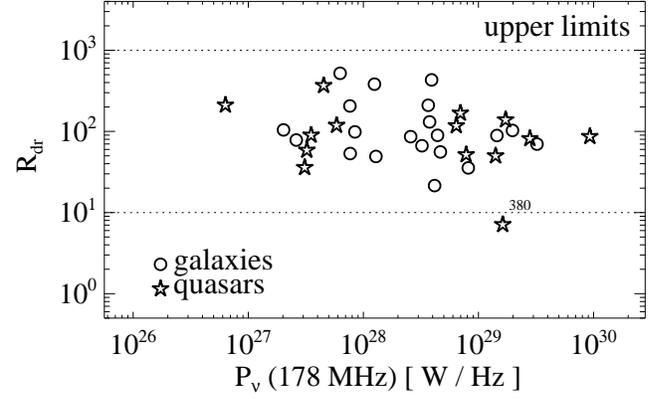,width=9.8cm,clip=true}
  \caption[]{ \label{msxxxx_fig_p80_to_p178_vs_p178_ul} 
    Same as Fig.\,\ref{msxxxx_fig_p80_to_p178_vs_z}, but for
    the sources from 
    Fig.\,\ref{msxxxx_fig_seds_3} with IR upper limits only.
  }
\end{figure}

With regard to the "matching strategy" applied by van
Bemmel et al. (2000) and Meisenheimer et al. (2001)
this means that, if we choose a {\it galaxy} -- {\it quasar} pair 
matching in 178 MHz radio lobe power, i.e. lying in a
vertical bin of Fig.\,\ref{msxxxx_fig_p80_to_p178_vs_z}, they
exhibit a similar R$_{\rm dr}$, exactly as predicted by the unified
schemes. Some exceptional cases are discussed below.
Also, if we plot R$_{\rm dr}$ versus redshift, or if we use the MIR power at 15 $\mu$m (with
values given in Table\,\ref{msxxxx_tab_luminosities}) instead of the
FIR power at
70 $\mu$m, all along the abscissa the R$_{\rm dr}$ distributions look
quite similar for {\it galaxies} and {\it quasars}. 

As illustrated in Fig.\,\ref{msxxxx_fig_p80_to_p178_vs_p178_ul}
with values given in Table\,\ref{msxxxx_table_rdr},
the FR\,2 NLRGs and quasars from Fig.\,\ref{msxxxx_fig_seds_3},
for which we obtained only upper limits for the dust luminosity,
populate the same range of R$_{\rm dr}$ as the detected sources from Fig.\,\ref{msxxxx_fig_seds} do.
Obviously  the upper limits are so high that their constraints are
consistent with our conclusions on
the {\it unification}. Future data with the Spitzer and HERSCHEL
satellites may consolidate this.

The range of R$_{\rm dr}$ values spans a factor of about ten
around the average R$_{\rm dr}$ $\sim$ 100 as 
indicated by the dotted lines in Fig.\,\ref{msxxxx_fig_p80_to_p178_vs_z}.
This appears large at first glance, but it becomes actually small with regard to the 
range of more than four decades of power.
In view of the complexity of the objects, with a diversity of the
strength of the thermal IR dust bump,  evolving giant radio lobes
(Kaiser et al. 1997, Blundell \& Rawlings 1999)
and various circumgalactic environments (Barthel \& Miley 1988),
such a dispersion is certainly not surprising.
Therefore, the striking similarity of the R$_{\rm dr}$ distributions
for {\it quasars} and {\it galaxies} provides strong
evidence in favour of the basic role of the orientation-dependent {\it unification}.

\subsubsection{Other influences}
\label{section_other_influences}

When looking at the radio maps, 
it becomes clear that the jet and lobe phenomena
are of course also influenced 
by the circumgalactic environment and by evolution.
Therefore it appears not  
appropriate, to consider and apply the three ``principles'' 
{\em unification}, {\em environment} and {\em evolution} 
on an exclusive either--or basis, rather the 
interplay of all three principles will help to 
understand the commonalities and differences of
{\it galaxies} and {\it quasars}.  The task now is to find
strategies for disentangling the effects of the three principles. 
Firstly, we discuss our data in view of the receding torus model --
this topic may be considered as {\em evolution}
of the torus with luminosity.

~

\hspace{-0.5cm}{\bf 4.2.3.1 Receding torus model}

~

In this modification of the general orientation-unifying model
(Lawrence 1991, Hill et al. 1996, review by Simpson 2003 and references therein)
both the {\it galaxies} and the {\it quasars} are basically surrounded by a dusty torus, but  
the inner wall of the torus recedes with increasing luminosity of the central engine
due to dust sublimation (r$_{\rm sub}$ $\propto$ L$^{\rm 0.5}$, e.g. Netzer \& Laor 1993). 
Under the assumption, that the scale height of the dust
torus is independent of luminosity, the opening angle of the bi-cone increases with luminosity.

When defining those objects with direct view to the broad emission line region
simply as {\it quasars} and all others as {\it galaxies},  
then the receding torus model predicts that the {\it quasar} fraction
increases with luminosity.
In fact, such a transition of the {\it quasar} fraction from about 20\% to higher values is observed
at a 151 MHz luminosity threshold of about 10$^{\rm 26.5}$ W/Hz (Willott et al. 2000).
However, in all the luminosity bins above 10$^{\rm 26.5}$ W/Hz the quasar fraction is 40\%
at the 1-$\sigma$ level (Willott et al. 2000).
This constancy of the quasar fraction at high luminosities is surprising in the frame of a
pure receding torus model.
It suggests that above a luminosity threshold -- possibly around 10$^{\rm 26.5}$ W/Hz --
the geometry of the receding torus does not change significantly and that the bi-cone
opening angle has reached a maximum value compatible with other constraints. 
All of our FR\,2 sources have 178 MHz luminosities higher than 10$^{\rm 26.5}$ W/Hz
(Fig.\,\ref{msxxxx_fig_p80_to_p178_vs_z}), hence lie above this threshold.
With regard to our 3CR data we emphasize two points:

1) If {\it galaxies} and  {\it quasars} belong to the same parent population, then --
  irrespective of whether a change of the source geometry with luminosity occurs or not --
  any orientation-dependent {\it unification} predicts that for a given luminosity bin
  the dust-to-lobe luminosity ratio R$_{\rm dr}$ should be
  the same for the {\it galaxy} and the {\it quasar}.
  If the {\it galaxy} had a lower R$_{\rm dr}$ than the {\it quasar},
  this would point to a weaker central engine in the {\it galaxy}.
  In such a case, however, we cannot determine from the current data, whether this
  difference between {\it galaxy} and {\it quasar} is accompanied by different
  torus properties -- and even if we could, this information would be only of moderate relevance
  with regard to other effects like lobe evolution and circumgalactic
  environment discussed in the next subsection. 
  Nevertheless, our 3CR data show a striking similarity of the R$_{\rm dr}$ distributions
  for the {\it quasars} and the {\it galaxies} (Fig.\,\ref{msxxxx_fig_p80_to_p178_vs_z})
  and any difference or decline with luminosity can
  be established at most at the 1-$\sigma$ level or less (Table\,\ref{msxxxx_table_rdr}).

  2) If the bi-cone opening angle increases significantly
  with luminosity, then 
  the covering angle of the dust, which absorbs the radiation from the
  central engine, decreases. Then one would expect that -- relative 
  to another measure of the AGN strength -- the IR luminosity declines.
  For example an increase of the half-opening angle of the bi-cone from
  30$\degr$ to 60-65$\degr$ results in a decrease of the dusty
  solid angle seen from the central source
  by a factor of about two (cos30$\degr$/cos65$\degr$=2.05).
  We searched for such a relative decline of the IR luminosity in two data samples:
  
  \begin{itemize}
  \item [a)] 
    If in {\it optically selected quasars}
    the blue continuum from the central engine
    is seen largely free of extinction, then the ratio of blue to IR luminosity
    should  decline.
    We examined the sample of 64 Palomar Green quasars with good near- and mid-IR data
    and covering four decades in luminosity (Fig.\,4 and Table\,2 in Haas et al. 2003).
    We find that for all bins above L$_{\rm B}$=10$^{\rm 11}$ L$_{\odot}$ (corresponding to
    M$_{\rm B}$$<$-24 mag)
    the mean ratios of L$_{\rm NIR}$/L$_{\rm B}$ and L$_{\rm MIR}$/L$_{\rm B}$ are clearly constant
    at the 1-$\sigma$ level with values about 1.1 $\pm$ 0.5. But
    for the low luminosity bins with L$_{\rm B}$$<$10$^{\rm 11}$ L$_{\odot}$
    the ratios are higher by a factor of about 1.5 and 1.7, respectively,
    although the differences are only marginally significant.
    
  \item [b)]
    For the {\it radio-loud 3CR objects} one may expect a decline of R$_{\rm dr}$ with luminosity,
    but our data do not show a clear trend at a statistically significant level
    (Fig.\,\ref{msxxxx_fig_p80_to_p178_vs_z}, Table\,\ref{msxxxx_table_rdr}).
  \end{itemize}
  
  Thus we conclude that the effects of the receding torus with increasing bi-cone opening angle
  may be significant at lower luminosity levels for Seyfert galaxies and radio galaxies having 
  178 MHz luminosities below  10$^{\rm 26.5}$ W/Hz,
  but that at the high luminosity level of quasars
  the receding torus model in its originally proposed rigid form plays only a minor role
  and needs some further refinement.


One refinement should focus on the torus scale height: 
For the receding torus model the assumption of a constant torus scale height is most critical.
It may not be justified if the AGN phenomenon is accompanied by nuclear starbursts and
if the star formation rate (SFR) grows with AGN luminosity.
Then supernovae will blow-up the scale height $h$ of the torus
(and the disk) by injection of kinetic energy.
For example, three-dimensional numerical simulations of the interstellar matter located 
in the combined gravitational potential of both a central black hole and a stellar system 
yield $h$ $\propto$ SFR$^{\rm 0.5}$ $\propto$ L$^{\rm 0.5}$ (Wada \& Norman 2002).
If this is the case, then both $h$ and r$_{\rm sub}$ scale with L$^{\rm 0.5}$ 
and the opening angle of the bi-cone does not change with luminosity.
This refinement would explain (1) that the quasar fraction
remains constant at 151 MHz luminosities above 10$^{\rm 26.5}$ W/Hz, and (2) that the ratios
L$_{\rm NIR}$/L$_{\rm B}$  and L$_{\rm MIR}$/L$_{\rm B}$ do not decline
at luminosities above L$_{\rm B}$=10$^{\rm 11}$ L$_{\odot}$ (M$_{\rm B}$$<$-24 mag).

~

\hspace{-0.5cm}{\bf 4.2.3.2 Environment and lobe evolution}

~

The ISO data suggest, that the FIR and MIR 
luminosities actually yield good estimates for the strength of the
AGN. 
Now, under the premise that the {\em unification} is basically 
valid, sensitive infrared measurements offer to pursue the following
strategy proposed by Barthel \& Arnaud (1996):
We adopt the IR luminosity as primary measure for the AGN power and
examine the strength of the radio lobes. Then 
deviations in the dust-to-lobe luminosity ratio allow,
free of orientation bias, to explore the additional 
influence of environment and evolution onto the extended radio lobes.  
In the following, we consider the extreme cases of high and low  R$_{\rm dr}$
values in Fig.\,\ref{msxxxx_fig_p80_to_p178_vs_z}.

High values of R$_{\rm dr}$ are expected for 
sources, which are either rather dust-rich or do not show strong radio
lobe activity.
Examples are  the "cool" FR\,2 galaxies 3C\,321 and 
3C\,459 both with a high FIR contribution possibly enhanced by starbursts.  
Very extreme examples for high R$_{\rm dr}$ sources
are radio-quiet quasars lying even above the upper plot range
of Fig.\,\ref{msxxxx_fig_p80_to_p178_vs_z}.

Low values of R$_{\rm dr}$ are expected for sources,
which are either rather dust-poor or do show exceptional radio
lobe activity - possibly caused by a special circumgalactic
environment or by a high evolutionary state of the lobes.
For example, 3C\,405 (Cygnus A) has exceptionally 
bright radio lobes (Barthel \& Arnaud, 1996) and shows a low R$_{\rm dr}$ $<$ 10, 
confirming the suggestion by Barthel \& Arnaud, that radio 
luminosity is not generally a good measure of AGN power. 
The ISO data show that the dust emission may serve as a better isotropic 
indicator for the AGN strength. 
Also the galaxy 3C\,295 lies at the low end of 
the FIR-to-radio distribution, indicating high efficiency in creating the 
radio lobe luminosity. 
Note that 3C\,295 is located in a dense X-ray bright cluster, 
suggesting the influence of the environment in enhancing the 
lobe power.
Also 3C\,380 is peculiar in that it shows both 
a relatively flat radio spectrum probably seen pole-on and 
bright FR\,2 lobes. A possible way out of this dilemma 
is that there is an intrinsic bend in the jet
(Wilkinson et al. 1991). In this case a strong interaction with the 
circumgalactic medium may provide an extraordinarily high radio 
luminosity, too.

The extreme cases with high and low  R$_{\rm dr}$ show that  
the dispersion of the R$_{\rm dr}$ distribution in general
may be well understood in terms of  
evolution of the radio lobes and by effects of the circumgalactic environment.

If evolutionary and environmental effects are superposed
on the pure orientation-dependent unification,
then it is a priori not clear whether the increased 
dispersion affects {\it galaxies} and
{\it quasars} in the same manner, or whether a bias
is introduced. 
A detailed investigation of the possible selection effects
(Chapter 3.5.3 in Haas 2001) shows
that there is a bias,
which limits the strategies to check the {\it unification}:
a radio source, which has already been recognised as a {\it quasar}
due to some criteria, will on average 
always exhibit a higher IR luminosity
than a source with matching radio luminosity, which has not yet
been identified as a {\it quasar}. 
Larger data samples are required in order to overcome this "strategical bias".

\section{Conclusions}
\label{section_conclusions}

We have obtained sensitive Infrared SEDs for a large sample of
3CR radio galaxies and quasars.
In order to check the orientation-dependent {\it unification} of powerful FR\,2 sources 
we considered 17 {\it galaxies} and 18
{\it quasars} with good IR detections. 
The main conclusions for this sample are:
\begin{itemize}
  \item[1) ]
    Both the {\it quasars} and the {\it galaxies}
    exhibit a high mid- to far-IR luminosity ratio
    typical for AGNs. 

  \item[2) ]
    The ratio of IR dust to radio lobe luminosity
    R$_{\rm dr}$ is similar for the {\it quasars} and the {\it galaxies}.
    Therefore, this result provides strong evidence in favour of the
    orientation-dependent {\it unification} of luminous
    (L$_{\rm IR}$ $\ga$ 10$^{\rm 11}$\,L$_{\odot}$)
    narrow line FR\,2 radio {\it galaxies} with {\it quasars}.

  \item[3) ]
    The dispersion of the R$_{\rm dr}$ distribution might be caused by 
    evolution by the radio lobes as well as effects of the circumgalactic
    environment.
 
  \item[4) ] The receding torus model in its original formulation
    is not valid at the high luminosity of our sources
    (L$_{\rm IR}$ $\ga$ 10$^{\rm 11}$\,L$_{\odot}$).
    Therefore we propose as a refinement that the scale height
    of the torus increases also with luminosity due to supernovae
    from starbursts accompanying the AGN.
    
    \end{itemize}
Compared with the detections,
the upper limits of the remaining  20 {\it galaxies} and 15
{\it quasars} are so high that they do not provide any deep
constraints or conflicts with the conclusions above.

\begin{table*}
 \caption[]{ Measured flux densities in mJy as a function of wavelength in $\mu$m.
Numbers in bold are detections above the 3-$\sigma$ level, the uncertainties are 10--30\%. 
Other numbers represent 3-$\sigma$ upper limits. 
The apertures used for each filter are listed underneath the wavelength; 
for 120-200\,$\mu$m the apertures were 90'' in the case of maps and 180'' in the case of chopped photometry. 
"m" and "c" denote mapped/chopped observations. The previous
publications of ISOPHOT flux values 
refer to: A = Andreani  et al. (2002), 
vB = van Bemmel et al. (2000),
F = Fanti et al. (2000), 
H1 = Haas et al. (1998), 
H2 = Haas et al. (2003), 
M = Meisenheimer et al. (2001), 
P = Polletta et al. (2000).
In addition to our photometry, we list also the 
SCUBA F$_{\rm 450\mu m}$ and F$_{\rm 850\mu m}$ literature values shown in the SED plots: They are taken from  
A = Archibald et al. (2001),
B = Best et al. (1998),
H = Hughes et al. (1993, at 800$\mu$m), 
P = Polletta et al. (2000),
R = Robson et al. (1998),
vB = van Bemmel et al. (2000),
W = Willott et al. (2002).
\label{msxxxx_tab_fluxes} 
}
\begin{center}
\scriptsize
\begin{tabular}{@{\hspace{1.0mm}}r|@{\hspace{1.0mm}}c@{\hspace{1.0mm}}|@{\hspace{1.0mm}}c@{\hspace{1.0mm}}r@{\hspace{1.0mm}}r@{\hspace{1.0mm}}r@{\hspace{1.0mm}}r@{\hspace{1.0mm}}r@{\hspace{1.0mm}}r@{\hspace{1.0mm}}r@{\hspace{1.0mm}}r|@{\hspace{1.0mm}}c@{\hspace{1.0mm}}r@{\hspace{1.0mm}}r@{\hspace{1.0mm}}r@{\hspace{1.0mm}}r|@{\hspace{1.0mm}}c@{\hspace{1.0mm}}r@{\hspace{1.0mm}}r@{\hspace{1.0mm}}r@{\hspace{1.0mm}}r@{\hspace{1.0mm}}r|r@{\hspace{1.0mm}}r@{\hspace{1.0mm}}r}
3C   &publ&map/ &  4.8 & 7.3 & 10 & 11.5 & 12.8 & 16 & 20 & 25 &map/ & 60 & 80 & 90 & 100 &map/ & 120 & 150 & 170 & 180 & 200 & 450 & 850 & 1200 \\
     &    &chop
     &  23''& 23''&23'' & 23''& 23''& 23''&52''&52''  &chop& 45''&45''&45''&45''&chop&     &     &     &     &     &  8''& 15''&  11''\\
 &   & &  & & & & & & &                               &&      & &  &     && & & & & &&&\\
         \hline
 &   & &  & & & & & & &                               &&      & &  &     && & & & & &&&\\
  2.0\parbox{0cm}{$^{\rm *}$}&  &c&  & & & & & & & $<$21   &m& {\bf 72}  & & & {\bf 158} &m& & & $<$609 & & &&&\\ 
  9.0& M &c&  {\bf 3} & & & & $<$42 & & {\bf 50} &         &c& $<$144    & & & $<$246    &c& & & $<$327 & & &&$<$11.4&\\ 
 16.0& F &c&  & & & & & & &                                &c& $<$129    & & $<$162&     &c& & & $<$177 & & $<$450 &&&$<$3.3\\ 
 17.0\parbox{0cm}{$^{\rm *}$}&  &m&  & & & $<$48 & & & &   & &           &        & &    &m& & & $<$33 & & &&&{\bf 290.0}\\ 
 19.0& vB,F & &  & & & & & & &                              &m& $<$60     & & $<$42 &     &m& & & $<$117 & & $<$495 &&&\\ 
 &   & &  & & & & & & &                               &&      & &  &     && & & & & &&&\\
 20.0& H1,M &m&  & & & & & & & $<$33                       &c& {\bf 97}  & & & {\bf 223} &m& & & {\bf 167} & & &&&$<$5.7\\ 
 33.0&      & &  & & & & & & &                             &m&&&{\bf 161}& & &           & & & & &&&{\bf 9.5}\\ 
 33.1&  &m&  & & & {\bf 14} & & & & {\bf 26}               &m& {\bf 48}  & & {\bf 41} &  &m& & & $<$66 & & &&&{\bf 5.8}\\ 
 34.0& F & &  & & & & & & &                                &m& $<$48     & & $<$60    &  &m& & & $<$189 & & $<$426 &&&$<$6.0\\ 
 42.0& vB,F & &  & & & & & & &                              &m& $<$78     & & $<$45    &  &m& & & $<$417 & & $<$360 &&&{\bf 5.2}\\ 
 &   & &  & & & & & & &                               &&      & &  &     && & & & & &&&\\
 46.0& A,F & &  & & & & & & &                              &c& $<$48     & & $<$75    &  &c& & & $<$264 & & $<$900 &&&\\ 
 47.0& M,P &m&  $<$9 & & & {\bf 10} & & & & {\bf 43}       &c& {\bf 103} & & & {\bf 159} &m& & & {\bf 164} & & &&&{\bf 10.8}\\ 
 48.0& H1,M &c&  {\bf 20} & & & & {\bf 50} & & {\bf 108} & &c& {\bf 460} & & & {\bf 829} &m& {\bf 730} & & {\bf 554} & {\bf 420} & &$<$63&{\bf  7.2}&\\ 
 49.0&  & &  & & & & & & &                                 &c& {\bf 120} & & $<$162 &    &c& & & $<$585 & & $<$675 &&&\\ 
 61.1&  &m&  & & & $<$3 & & & & $<$15                      &m& {\bf 33}  & & {\bf 44} &  &m& & & $<$60 & & &&&{\bf 3.0}\\ 
 &   & &  & & & & & & &                               &&      & &  &     && & & & & &&&\\
 65.0& M &m&  {\bf 2} & & & {\bf 8} & & & & {\bf 26}       &c& {\bf 129} & & & $<$129    &m& & & $<$99 & & &&$<$11.4\parbox{0cm}{$^{\rm H}$}&\\ 
 67.0& vB,F &c&  & & & $<$15 & & & &                        &m& $<$78     & & $<$60 &     &m& & & $<$444 & & &&&\\ 
 68.1&  &c&  & & & & & $<$39 & $<$57 & $<$54               & &           & & &           &c& & & & & &&&$<$6.3\\ 
 79.0& F & &  & & & & & & &                                &c& {\bf 144} & & {\bf 116} & &c& & & $<$585 & & $<$900 &&&$<$2.1\\ 
 84.0&   & &  & & & & & & &                                & &           & & &           &m& & & {\bf 4153} & & {\bf 1956} &&&\\ 
 &   & &  & & & & & & &                               &&      & &  &     && & & & & &&&\\
109.0&  &m&  & {\bf 31} & & {\bf 58} & & & & {\bf 170}     &m& {\bf 203} & & & {\bf 158} &m& & $<$582 & & & $<$813 &&&{\bf 39.0}\\ 
138.0& M &m&  & & & {\bf 11} & & & & {\bf 24}              &m& $<$48     & & & $<$105    &m& & & $<$117 & & &&{\bf 107}&\\ 
234.0& &c& {\bf 16} & {\bf 41} & & {\bf 100}&&&& {\bf 276} &c& {\bf 315} & & & {\bf 261} &c& & $<$162 & & & $<$1125 &&&{\bf 4.1}\\ 
245.0&  & &  & & & & & & &                                 &c& {\bf 196} & & & $<$153    &c& & & $<$558 & & &&&\\ 
249.1& &c& {\bf 9} & {\bf 14} & & {\bf 18} & & & &{\bf 40} &m& {\bf 60}  & & & {\bf 40}  &m& & $<$90 & & & $<$90 &&$<$7.2\parbox{0cm}{$^{\rm P}$}&\\ 
 &   & &  & & & & & & &                               &&      & &  &     && & & & & &&&\\
268.3& A,F &c&  & & & $<$24 & & & &                        &c& $<$69     & & $<$84 &     &c& & & $<$720 & & $<$750 &&&\\ 
268.4&  A& &  & & & & & & &                    &c& & & {\bf 54}\parbox{0cm}{$^{\rm c}$}& &c& & &{\bf 586}\parbox{0cm}{$^{\rm c}$}&&&$<$37.5\parbox{0cm}{$^{\rm W}$}&5.1\parbox{0cm}{$^{\rm W}$}&\\ 
270.0\parbox{0cm}{$^{\rm *}$}& & & & & & & & & &           &c& $<$291    & & {\bf 492} & &c& & & {\bf 661} & & $<$900 & {\bf 200}& {\bf 210}&  \\ 
272.1&  & &  & & & & & & &                                 &c& {\bf 700} & & {\bf 1000}& &c& & & {\bf 703} & & $<$900 &&&{\bf 137.0}\\ 
273.0& M & &  &&&& & & &                           &c& {\bf 1124}&{\bf 1291}&&{\bf 1348} &c& {\bf 1546}&{\bf 1113}&{\bf 1292}&{\bf 1056}&{\bf 1091} &&&\\ 
 &   & &  & & & & & & &                               &&      & &  &     && & & & & &&&\\
274.0&  & &  & & & & & & &                                 &c& {\bf 610} & & {\bf 954} & &c& & & {\bf 322} & & $<$516 &{\bf 1320}&{\bf 2480}&\\ 
274.1& F & &  & & & & & & &                                &c& $<$222    & & $<$432    & &c& & & $<$702 & & $<$675 &&&$<$5.1\\ 
277.1\parbox{0cm}{$^{\rm *}$}& vB  & &  & & & & & & &       &m& $<$144    & & $<$90     & &m& & & $<$93 & & &&&\\ 
277.2& F & &  & & & & & & &                                &m& {\bf 146} & & {\bf 92}  & &m& & & $<$156 & & &$<$32\parbox{0cm}{$^{\rm A}$}&$<$3.0\parbox{0cm}{$^{\rm A}$}&\\ 
280.0& A,M &c&  $<$3 & $<$12 & & & $<$66 & & $<$96       & &c& {\bf 145} & &$<$126&$<$189&c& & & $<$231 & & &&&\\ 
 &   & &  & & & & & & &                               &&      & &  &     && & & & & &&&\\
280.1& & & & & & & & & &                                   &c&           & & $<$75     & &c& & & & &  &$<$71.4\parbox{0cm}{$^{\rm W}$}&$<$5.2\parbox{0cm}{$^{\rm W}$}&\\ 
284.0\parbox{0cm}{$^{\rm *}$}& F & & & & & & & & &         &c&       & {\bf 157}&$<$285& &c& & & $<$135 & & $<$567 &&& \\ 
286.0& A,M &c&  $<$9 & & & & $<$93 & & $<$117 &            &c& {\bf 131} & & & {\bf 105} &c& & & $<$186 & & &&{\bf 259}&\\ 
287.0& A,M &c&  $<$18 & & & & $<$69 & & $<$126 &           &c& $<$156    & & & $<$78     &c& & & & & &&{\bf 85}&\\ 
288.1\parbox{0cm}{$^{\rm *}$}& A &c&  & & & & & & & $<$18  &c& $<$75     & & & {\bf 127} &c& & & $<$105 & & &&&$<$4.2\\ 
 &   & &  & & & & & & &                               &&      & &  &     && & & & & &&&\\
289.0&  & &  & & & & & & &                                 &m&           & & $<$39     & &m& & & $<$48 & & &&&\\ 
293.0&  & &  & & & & & & &                                 & &           & & &           & & & &       & & &&&$<$4.2\\ 
295.0& A,F,M &c&  {\bf 3} & $<$12 & & & $<$30 & & $<$48 &  &c& {\bf 87}  & & & $<$291    &c& & & $<$111    & &     & & & \\ 
298.0\parbox{0cm}{$^{\rm *}$}&M&c&$<$12&&&&{\bf 23}&&{\bf 29}&&c& {\bf 184}& & &{\bf 213}&c& & & {\bf 243} & &     &$<$71\parbox{0cm}{$^{\rm W}$}&21.1\parbox{0cm}{$^{\rm W}$}& \\ 
299.0& F& &  & & & & & & &                                 & &           & & &           &c& & & &           & $<$450& & & \\
 &   & &  & & & & & & &                               &&      & &  &     && & & & & &&&\\
303.1& F & &  & & & & & & &                                &c& $<$111    & & {\bf 74}  & &c& & & $<$480 & & $<$390 &&&\\ 
309.1& A,F & &  & & & & & & &                              & & {\bf 118} & & & $<$144    &c& & & $<$273 & & &&$<$26&\\ 
318.0& F & &  & & & & & & &                                &c& {\bf 157} & & {\bf 284} & &c& & & {\bf 391} & & $<$657 &$<$32\parbox{0cm}{$^{\rm W}$}&7.8\parbox{0cm}{$^{\rm W}$}&\\ 
321.0&  & &  & & & & & & &                                 &c&           & & {\bf 780} & &m& & & {\bf 402} & & &&$<$20&\\ 
323.1\parbox{0cm}{$^{\rm *}$}&vB,H2&c&{\bf 11}&{\bf 16}&&{\bf 25}&&&&{\bf 43}&m&{\bf 45}&&{\bf 33} & &m& & & $<$57 & & &&$<$9.0\parbox{0cm}{$^{\rm vB}$}&\\ 
 &   & &  & & & & & & &                               &&      & &  &     && & & & & &&&\\
324.0& M & &  & & & & & & &                                &c& $<$102    & & & $<$231    &c& & & $<$237 & & &$<$21\parbox{0cm}{$^{\rm B}$}&3.0\parbox{0cm}{$^{\rm B}$}&\\ 
325.0& A,M &c&  $<$39 & $<$18 & & & $<$75 & & $<$81&       &c& {\bf 160} & & & {\bf 140} &c& & & $<$126 & & &&$<$25&\\ 
330.0& F & &  & & & & & & &                                &m& $<$75     & & $<$45 &     &m& & & $<$123 & & $<$540 &&&$<$3.2\\ 
332.0\parbox{0cm}{$^{\rm *}$}& &m &  & & & & & & & $<$150  &m& $<$216    & & $<$132 &    &m& & & $<$225 & & &&&$<$10.2\\ 
334.0& vB  & &  & & & & & & &                               &m& {\bf 86}  & & {\bf 56} &  &m& & & $<$60 & & &&$<$14&{\bf 8.3}\\ 
 &   & &  & & & & & & &                               &&      & &  &     && & & & & &&&\\
\hline
\end{tabular}
\end{center} 
$^{*}$ contained in Spinrad et al. (1985), but not in Laing et al. (1983);
$^{c}$ contaminated by NGC\,4138; 
$^{x}$ possibly contaminated by an asteroid;\\
$^{s}$ integrated over spectral continuum obtained with ISOPHOT-S; 
$^{+}$ also: 1060 mJy at 65 $\mu$m
\end{table*}

\addtocounter{table}{-1}
\begin{table*}
 \caption[]{ continued. 
\label{msxxxx_tab_fluxes_a} 
}
\begin{center}
\scriptsize
\begin{tabular}{@{\hspace{1.0mm}}r|@{\hspace{1.0mm}}c@{\hspace{1.0mm}}|@{\hspace{1.0mm}}c@{\hspace{1.0mm}}r@{\hspace{1.0mm}}r@{\hspace{1.0mm}}r@{\hspace{1.0mm}}r@{\hspace{1.0mm}}r@{\hspace{1.0mm}}r@{\hspace{1.0mm}}r@{\hspace{1.0mm}}r|@{\hspace{1.0mm}}c@{\hspace{1.0mm}}r@{\hspace{1.0mm}}r@{\hspace{1.0mm}}r@{\hspace{1.0mm}}r|@{\hspace{1.0mm}}c@{\hspace{1.0mm}}r@{\hspace{1.0mm}}r@{\hspace{1.0mm}}r@{\hspace{1.0mm}}r@{\hspace{1.0mm}}r|r@{\hspace{1.0mm}}r@{\hspace{1.0mm}}r}
3C   &publ&map/ &  4.8 & 7.3 & 10 & 11.5 & 12.8 & 16 & 20 & 25 &map/ & 60 & 80 & 90 & 100 &map/ & 120 & 150 & 170 & 180 & 200 & 450 & 850 & 1200 \\
     &    &chop
     &  23''& 23''&23'' & 23''& 23''& 23''&52''&52''  &chop& 45''&45''&45''&45''&chop&     &     &     &     &     &  8''& 15''&  11''\\
 &   & &  & & & & & & &                               &&      & &  &     && & & & & &&&\\
         \hline
 &   & &  & & & & & & &                               &&      & &  &     && & & & & &&&\\
336.0& &c& $<$15 &$<$6 & $<$12 & & & $<$36 & $<$75 & $<$42 &c& $<$96     & & & $<$168    &c& & & $<$468 & & &$<$47&$<$5.5&\\ 
337.0& A,F & &  & & & & & & &                              &m& $<$63     & & $<$78 &     &m& & & $<$102 & & &&&\\ 
340.0&  & &  & & & & & & &                                 &m&           & & $<$99 &     &m& & & $<$168 & & &$<$27\parbox{0cm}{$^{\rm A}$}&$<$2.6\parbox{0cm}{$^{\rm A}$}&\\ 
343.1& A,F & &  & & & & & & &                              &c& $<$108    & & & $<$462    &c& & & $<$411 & & &&&{\bf 7.6}\\ 
351.0& vB,H2 &c& {\bf 18}&{\bf 31}&&{\bf 50}&&&&{\bf 112}&m&{\bf 201}&&{\bf 168}&{\bf 151}&m& & {\bf 94} & {\bf 66} & & {\bf 48} &&&$<$3.3\\ 
 &   & &  & & & & & & &                               &&      & &  &     && & & & & &&&\\
352.0& A & &  & & & & & & &                                &m&           & & $<$75 &     &m& & & $<$156 & & &&&\\ 
356.0& A & &  & & & & & & &                                &m&           & & {\bf 27} &  &m& & & $<$84 & & &$<$60\parbox{0cm}{$^{\rm A}$}&$<$3.1\parbox{0cm}{$^{\rm A}$}&\\ 
368.0& M &c&  {\bf 4}\parbox{0cm}{$^{\rm x}$} & {\bf 16}\parbox{0cm}{$^{\rm x}$}  & & & $<$120 & & $<$162& &c& {\bf 94}  & & & $<$144    &c& & & $<$168 & & &$<$45\parbox{0cm}{$^{\rm A}$}&4.1\parbox{0cm}{$^{\rm A}$}&\\ 
371.0\parbox{0cm}{$^{\rm *}$}& &      &  & & & & & & &     &c&           & & {\bf 397} & &c& & & {\bf 275} & & &&&\\ 
var.                             &  & &  &&&&&&&           &c&           & & {\bf226}  & &c& & & {\bf 215} & & &&&\\
 &   & &  & & & & & & &                               &&      & &  &     && & & & & &&&\\
380.0& M &c&  $<$9 & {\bf 7} & & $<$18 & & & & $<$66       &m& {\bf 54}  & & & {\bf 56}  &m& & & {\bf 83} & & &&{\bf 659}&\\ 
381.0&  &m&  & & & {\bf 19} & & & & {\bf 52}               &m& {\bf 74}  & & $<$78 &     &m& & & {\bf 42} & & &&&$<$2.1\\ 
390.3& &c&&{\bf 55}\parbox{0cm}{$^{\rm s}$}&&&&&&{\bf 139} &c& {\bf 191} & & $<$1233 &   &c& $<$963 & & & $<$984 & &$<$385&$<$14&\\ 
401.0& F & &  & & & & & & &                                &c& $<$78     & & $<$99 &     &c& & & $<$351 & & $<$270 &&&{\bf 12.2}\\ 
405.0\parbox{0cm}{$^{\rm *}$}&H1,M,P&c&$<$12&{\bf 40}&&&{\bf250}&&{\bf 590} &&c&{\bf 3034} &&&{\bf 2155} &m& {\bf 994} & & &
     {\bf 419} & &$\sim$230\parbox{0cm}{$^{\rm R}$}&$\sim$490\parbox{0cm}{$^{\rm R}$}&{\bf 628}\\ 
 &   & &  & & & & & & &                               &&      & &  &     && & & & & &&&\\
427.1& M &c&  $<$48 & $<$15 & & & $<$114 & & $<$75 &       &c& $<$135    & & & $<$204    &c& & & $<$180 & & &&&\\ 
437.0& M &c&  $<$21 & $<$12 & & & $<$102 & & $<$138 &      &c& $<$126    & & & $<$246    &c& & & $<$189 & & &$<$52\parbox{0cm}{$^{\rm W}$}&$<$3.0\parbox{0cm}{$^{\rm W}$}&\\ 
441.0& F & &  & & & & & & &                                &c& $<$174    & & $<$153 &    &c& & & $<$234 & & $<$324 &&&$<$3.0\\ 
445.0\parbox{0cm}{$^{\rm *}$}&&c&{\bf 46}&{\bf 73}&&&&&&{\bf 360}&c&{\bf 435}&&& $<$357  &c& & $<$849 & & & $<$1647 &&&$<$4.8\\ 
454.3&  &c&  $<$12 & {\bf 16} & & & {\bf 34} & & $<$291 &  &c& {\bf 706} & & & {\bf 324} &m&{\bf 187}&&& {\bf 248} & &{\bf 1370}&{\bf 2780}&{\bf3716.0}\\ 
 &   & &  & & & & & & &                               &&      & &  &     && & & & & &&&\\
459.0\parbox{0cm}{$^{\rm *}$}&F&c&&&&&&&&$<$450 &c& {\bf 1035}\parbox{0cm}{$^{\rm +}$}&&{\bf 1375} & &c& & {\bf 1110}  & {\bf 815} & & $<$1413 &&&\\ 
460.0& vB  & &  & & & & & & &                               &m& $<$57     & & $<$54 &     &m& & & $<$351 & & &&&\\
 &   & &  & & & & & & &                               &&      & &  &     && & & & & &&&\\
\hline
\end{tabular}
\end{center} 
\end{table*}

\begin{table*}
 \caption[] {Types, redshifts and luminosity distances, restframe luminosities, dust temperatures and masses. 
The luminosity distance D$_{\rm L}$ is calculated adopting a $\Lambda$ cosmology with 
H$_0$ = 71 km\,s$^{-1}$\, Mpc$^{-1}$, $\Omega_{{\rm matter}}$ = 0.27
and $\Omega_{\Lambda}$ = 0.73.
The luminosities L$_{\rm opt}$ (0.5-1$\,\mu$m), L$_{\rm NIR}$ (3-10$\,\mu$m), 
L$_{\rm MIR}$ (10-40$\,\mu$m) and L$_{\rm FIR}$ (40-1000$\,\mu$m) 
are computed in the restframes of the objects from the ``envelopes'' 
of the SEDs. The luminosities give the total
emission from host galaxy and the core. 
The IR luminosities largely represent the thermal dust luminosities, 
except for some sources (like 3C\,371 or 3C\,380), which do not exhibit any prominent thermal bump 
and are excluded from the discussion.
The thermal dust emission power ${\nu}$P$_{\rm \nu}$ has been computed at restframe wavelengths 15 and 70 $\mu$m, respectively. 
The 178 MHz radio lobe power is computed from the observed 178 MHz fluxes, 
applying a K--correction with a spectral index $\alpha$ = 0.5. 
The choice of $\alpha$ has only little impact on the resulting lobe 
power and the relations discussed. R$_{\rm dr}$ is the ratio of dust power
to radio lobe power at 15 and 70 $\mu$m, respectively. 
The values of
T$_{\rm p}$ refer to the peak blackbody shown in 
the SED plots, the uncertainty is 
5-10\,K and can raise to 50\% in cases of poor FIR upper limits. 
The dust mass M$_{\rm dust}$ is estimated  according to Formula 6 
in Stickel et al. (2000) following Hildebrand (1983), 
with uncertainties up to an order of magnitude. 
A ``$\sim$'' marks highly uncertain L$_{\rm MIR}$/L$_{\rm FIR}$,  T$_{\rm dust}$ and M$_{\rm dust}$
values due to poor dust bump determinations or upper limit constraints, and these sources are not used in the discussion.
\label{msxxxx_tab_luminosities} 
} 
\scriptsize
\begin{center}
\begin{tabular}{@{\hspace{1.1mm}}r@{\hspace{1.1mm}}|l@{\hspace{1.1mm}}l@{\hspace{1.1mm}}l|r@{\hspace{1.1mm}}r|r@{\hspace{1.1mm}}r@{\hspace{1.1mm}}r@{\hspace{1.1mm}}r@{\hspace{2.1mm}}r|r@{\hspace{1.1mm}}r@{\hspace{1.1mm}}r@{\hspace{1.1mm}}r@{\hspace{1.1mm}}r|r@{\hspace{1.1mm}}r} 
3CR & & type\parbox{0cm}{$^{\rm *}$} & & redshift & D$_{\rm L}$ & L$_{\rm opt}$ & L$_{\rm NIR}$ & L$_{\rm MIR}$  & L$_{\rm FIR}$& L$_{\rm MIR}$/  & P(15 $\mu$m) & P(70 $\mu$m) & P(178 MHz) & R$_{\rm dr}$ & R$_{\rm dr}$ & T$_{\rm p}$ & M$_{\rm d}$ \\ 
 & & & &   &{ Mpc}   &
{ log\,[L$_{\odot}$] }              &
{ log\,[L$_{\odot}$] }              &
{ log\,[L$_{\odot}$] }              &
{ log\,[L$_{\odot}$] }              &
 L$_{\rm FIR}$ &
{ log\,[W] }              &
{ log\,[W] }              &
{ log\,[W] }              &
{ 15 $\mu$m }&
{ 70 $\mu$m }&
{ [K] }              &
{ log\,[M$_{\odot}$] }               \\
  & &   &&& &&&&&   &&   && && &          \\
\hline
  & &   &&& &&&& &  &&   && && &          \\
  2.0& QSR&   C&CSS&1.037& 6943&11.39&$<$12.00&   12.80&   12.61&       1.54&   39.11&   39.09&37.34&     59&     57&        55&        7.92 \\
  9.0& QSR&   L&   &2.012&15852&12.76&   13.85&$<$13.87&$<$13.62&$\sim$ 1.79&$<$40.28&$<$40.16&38.22&$<$ 115&$<$  87&$\sim$  50&$\sim$  8.99 \\
 16.0&LERG&FR\,2&   &0.405& 2193&10.38&$<$11.31&$<$11.85&$<$11.90&$\sim$ 0.89&$<$38.11&$<$38.45&36.13&$<$  95&$<$ 206&$\sim$  40&$\sim$  7.81 \\
 17.0&BLRG&   ?&   &0.220& 1081&10.37&   10.67&$<$10.65&$<$10.96&$\sim$ 0.49&$<$37.06&$<$37.54&35.74&$<$  21&$<$  63&$\sim$  40&$\sim$  6.91 \\
 19.0&NLRG&FR\,2&   &0.482& 2701&10.80&$<$11.29&$<$11.50&$<$11.47&$\sim$ 1.06&$<$37.87&$<$38.05&36.36&$<$  32&$<$  49&$\sim$  40&$\sim$  7.41 \\
&&&&& &&&&&&&&&&&&\\
 20.0&NLRG&FR\,2&   &0.174&  830& 9.93&   10.52&   10.95&   11.18&       0.58&   37.21&   37.68&35.83&     24&     70&        35&        7.45 \\
 33.0&NLRG&FR\,2&   &0.059&  261&10.31&    9.93&$<$10.06&   10.14&       0.83&$<$36.45&   36.72&34.92&$<$  34&     63&        35&        6.40 \\
 33.1&BLRG&FR\,2&   &0.181&  869&10.15&   10.96&   11.00&   10.61&       2.47&   37.47&   37.14&35.36&    129&     60&        50&        6.08 \\
 34.0&NLRG&FR\,2&   &0.689& 4177&11.01&$<$11.70&$<$12.09&$<$12.03&$\sim$ 1.16&$<$38.41&$<$38.58&36.76&$<$  45&$<$  67&$\sim$  40&$\sim$  7.95 \\
 42.0&NLRG&FR\,2&   &0.395& 2128&10.60&$<$10.69&$<$11.12&$<$11.28&$\sim$ 0.70&$<$37.42&$<$37.86&36.14&$<$  19&$<$  53&$\sim$  40&$\sim$  7.23 \\
&&&&& &&&&&&&&&&&&\\
 46.0&NLRG&FR\,2&   &0.437& 2400&10.91&$<$11.37&$<$11.63&$<$11.61&$\sim$ 1.06&$<$37.98&$<$38.17&36.18&$<$  64&$<$  99&$\sim$  40&$\sim$  7.54 \\
 47.0& QSR&   L&   &0.425& 2322&11.09&   11.62&   12.01&   11.91&       1.24&   38.38&   38.48&36.56&     65&     84&        35&        8.18 \\
 48.0& QSR&   C&CSS&0.367& 1950&11.64&   12.21&   12.44&   12.47&       0.93&   38.79&   39.02&36.72&    117&    203&        35&        8.70 \\
 49.0&NLRG&FR\,2&CSS&0.621& 3675&10.48&$<$11.65&   12.28&   12.34&       0.88&   38.52&   38.90&36.58&     87&    209&        40&        8.26 \\
 61.1&NLRG&FR\,2&   &0.186&  895& 9.89&   10.02&   10.28&   10.59&       0.49&   36.47&   37.16&35.76&      5&     25&        40&        6.54 \\
&&&&& &&&&&&&&&&&&\\
 65.0&NLRG&FR\,2&   &1.176& 8125&11.11&   12.71&   13.10&   12.51&       3.94&   39.42&   38.95&37.50&     83&     29&        80&        7.17 \\
 67.0&BLRG&FR\,2&CSS&0.310& 1600&10.50&$<$10.96&$<$11.17&$<$11.17&$\sim$ 1.01&$<$37.54&$<$37.75&35.80&$<$  55&$<$  90&$\sim$  40&$\sim$  7.11 \\
 68.1& QSR&   L&   &1.238& 8664&12.13&$<$12.31&$<$12.65&$<$13.06&$\sim$ 0.39&$<$38.96&$<$39.63&37.49&$<$  30&$<$ 139&$\sim$  40&$\sim$  8.01 \\
 79.0&NLRG&FR\,2&   &0.256& 1283&10.66&   11.08&   11.58&   11.38&       1.60&   37.94&   37.94&36.08&     73&     73&        55&        6.63 \\
 84.0&NLRG& FR\,1&   &0.018&   74&10.61&   10.46&   10.87&   10.80&       1.18&   37.11&   37.30&33.85&   1831&   2779&        35&        7.01 \\
&&&&& &&&&&&&&&&&&\\
109.0&BLRG&FR\,2&   &0.306& 1577&11.28&   12.06&   12.22&   11.69&       3.38&   38.67&   38.21&36.12&    362&    126&        55&        6.95 \\
138.0& QSR&   C&CSS&0.759& 4707&11.45&   12.24&   12.38&$<$12.29&$\sim$ 1.23&   38.81&$<$38.86&37.14&     47&$<$  52&$\sim$  40&$\sim$  8.20 \\
234.0&NLRG&FR\,2&   &0.185&  890&10.56&   11.64&   11.95&   11.50&       2.81&   38.39&   38.04&35.76&    426&    189&        45&        7.17 \\
245.0& QSR&   L&   &1.029& 6876&11.83&   12.49&   13.09&   12.60&       3.03&   39.38&   38.94&37.32&    115&     42&        85&        7.08 \\
249.1& QSR&   L&   &0.311& 1607&11.60&   11.70&   11.66&   11.16&       3.14&   38.11&   37.57&35.83&    191&     55&        60&        6.26 \\
&&&&& &&&&&&&&&&&&\\
268.3&BLRG&FR\,2&CSS&0.371& 1975&10.27&$<$11.22&$<$11.57&$<$11.54&$\sim$ 1.08&$<$37.90&$<$38.09&36.02&$<$  77&$<$ 119&$\sim$  40&$\sim$  7.46 \\
270.0&NLRG& FR\,1&   &0.007&   31&10.28&    9.50&    8.84&    9.05&       0.61&   35.10&   35.51&33.03&$<$ 120&    305&        28&        5.81 \\
272.1&LERG& FR\,1&   &0.003&   12&10.25&    9.43&    8.13&    8.33&       0.63&   34.46&   34.77&31.90&    361&    736&        27&        5.13 \\
273.0& QSR&FR\,2&   &0.158&  747&12.13&   12.28&   12.13&   12.00&       1.35&   38.56&   38.47&35.91&    453&    368&        30&        8.53 \\
274.0&NLRG& FR\,1&   &0.004&   16&10.29&    9.56&    8.43&    8.50&       0.86&   34.72&   35.01&32.56&    144&    286&        35&        4.75 \\
&&&&& &&&&&&&&&&&&\\
274.1&NELR&FR\,2&   &0.422& 2303&10.67&$<$11.56&$<$12.12&$<$12.34&$\sim$ 0.60&$<$38.38&$<$38.93&36.35&$<$ 108&$<$ 382&$\sim$  40&$\sim$  8.30 \\
277.1& QSR&FR\,2&   &0.321& 1667&10.90&$<$11.23&$<$11.60&$<$11.16&$\sim$ 2.78&$<$37.93&$<$37.59&35.76&$<$ 147&$<$  67&$\sim$  40&$\sim$  6.95 \\
277.2&NLRG&FR\,2&   &0.766& 4761&10.89&$<$11.86&   12.71&   12.13&       3.82&   38.89&   38.51&36.89&    101&     42&        90&        6.58 \\
280.0&NLRG&FR\,2&   &0.996& 6602&10.98&$<$12.23&   12.97&   12.36&       4.02&   39.19&   38.73&37.49&     49&     17&        90&        6.81 \\
280.1& QSR&FR\,2&   &1.659&12481&11.95&$<$12.67&$<$13.03&$<$13.12&$\sim$ 0.82&$<$39.38&$<$39.61&37.70&$<$  48&$<$  82&$\sim$  40&$\sim$  8.28 \\
&&&&& &&&&&&&&&&&&\\
284.0&NLRG&FR\,2&   &0.239& 1187&10.82&$<$11.19&$<$11.49&   11.55&      0.89&$<$37.84&   38.11&35.58&$<$ 182&    344&        40&        7.49 \\
286.0& QSR&   C&CSS&0.849& 5410&11.94&   12.01&   12.66&   12.28&       2.42&   38.87&   38.71&37.33&     35&     24&        75&        7.00 \\
287.0& QSR&   C&CSS&1.055& 7094&11.92&$<$12.18&$<$12.51&$<$12.54&$\sim$ 0.93&$<$38.84&$<$39.10&37.40&$<$  28&$<$  50&$\sim$  40&$\sim$  8.47 \\
288.1& QSR&   L&   &0.961& 6313&11.75&   12.01&   12.66&   12.51&       1.43&   38.97&   39.02&37.03&     87&     98&        55&        7.78 \\
289.0&NLRG&FR\,2&   &0.967& 6362&11.21&$<$11.82&$<$12.16&$<$12.15&$\sim$ 1.02&$<$38.49&$<$38.71&37.16&$<$  21&$<$  36&$\sim$  40&$\sim$  8.08 \\
&&&&& &&&&&&&&&&&&\\
293.0&LERG& FR\,1&   &0.045&  197&10.52&   10.07&    9.98&   10.43&       0.36&   36.32&   36.85&34.04&    191&    655&        26&        7.41 \\
295.0&NLRG&FR\,2&   &0.461& 2559&11.09&   11.29&   11.81&   11.99&       0.67&   38.07&   38.57&37.15&      8&$<$  26&        40&        7.94 \\
298.0& QSR&FR\,2&   &1.436&10427&12.77&   13.32&   13.52&   13.19&       2.17&   39.88&   39.70&38.24&     44&     29&        40&        9.06 \\
299.0&NLRG&FR\,2&CSS&0.367& 1950&10.48&$<$11.05&$<$11.68&$<$12.35&$\sim$ 0.21&$<$37.78&$<$38.94&36.05&$<$  53&$<$ 780&$\sim$  40&$\sim$  8.30 \\
303.1&NLRG&FR\,2&CSS&0.267& 1347&10.61&   10.56&   10.83&   11.11&       0.52&   37.10&   37.71&35.55&     36&    145&        40&        7.07 \\
&&&&& &&&&&&&&&&&&\\
309.1& QSR&   C&CSS&0.905& 5858&12.05&   12.40&   12.79&   12.39&       2.51&   39.10&   38.82&37.36&     55&     29&        70&        7.22 \\
318.0& QSR&FR\,2&CSS&1.574&11690&11.62&$<$12.86&   13.56&   13.38&       1.51&   39.80&   39.89&37.76&    111&    134&        55&        8.64 \\
321.0&NLRG&FR\,2&   &0.096&  435&10.60&   10.60&   11.33&   11.36&       0.92&   37.45&   37.86&34.76&    497&   1258&        35&        7.49 \\
323.1& QSR&FR\,2&   &0.264& 1329&11.42&   11.62&   11.53&   10.88&       4.49&   38.03&   37.38&35.61&    261&     59&        60&        5.98 \\
324.0&NLRG&FR\,2&   &1.206& 8385&11.25&   12.12&$<$13.00&$<$13.01&$\sim$ 0.98&$<$39.16&$<$39.56&37.55&$<$  41&$<$ 103&$\sim$  50&$\sim$  8.42 \\
&&&&& &&&&&&&&&&&&\\
\hline
\end{tabular}
\end{center}
$^{\rm *}$ The types are abbreviated as:

QSR = quasar, 
BLRG = broad line radio galaxy, 
NLRG = narrow line radio galaxy, 
LERG = low-excitation radio galaxy, 

C = core dominated, 
L = lobe dominated, 
FR\,1, FR\,2 = Fanaroff-Riley classes 1 and 2, 

CSS = compact steep spectrum source, 
OVV = optically violent variable.

\end{table*}
\addtocounter{table}{-1}
\begin{table*}
 \caption[] {continued}
\scriptsize
\begin{center}
\begin{tabular}{@{\hspace{1.1mm}}r@{\hspace{1.1mm}}|l@{\hspace{1.1mm}}l@{\hspace{1.1mm}}l|r@{\hspace{1.1mm}}r|r@{\hspace{1.1mm}}r@{\hspace{1.1mm}}r@{\hspace{1.1mm}}r@{\hspace{2.1mm}}r|r@{\hspace{1.1mm}}r@{\hspace{1.1mm}}r@{\hspace{1.1mm}}r@{\hspace{1.1mm}}r|r@{\hspace{1.1mm}}r} 
3CR & & type\parbox{0cm}{$^{\rm *}$} & & redshift & D$_{\rm L}$ & L$_{\rm opt}$ & L$_{\rm NIR}$ & L$_{\rm MIR}$  & L$_{\rm FIR}$& L$_{\rm MIR}$/  & P(15 $\mu$m) & P(70 $\mu$m) & P(178 MHz) & R$_{\rm dr}$ & R$_{\rm dr}$ & T$_{\rm d}$ & M$_{\rm d}$ \\ 
 & & & &   &{ Mpc}   &
{ log\,[L$_{\odot}$] }              &
{ log\,[L$_{\odot}$] }              &
{ log\,[L$_{\odot}$] }              &
{ log\,[L$_{\odot}$] }              &
 L$_{\rm FIR}$ &
{ log\,[W] }              &
{ log\,[W] }              &
{ log\,[W] }              &
{ 15 $\mu$m }&
{ 70 $\mu$m }&
{ [K] }              &
{ log\,M$_{\odot}$ }               \\
  & &   &&& &&&&&   &&   && && &          \\
\hline
  & &   &&& &&&& &  &&   && && &          \\
325.0& QSR&   L&   &0.860& 5498&11.43&$<$13.10&   13.07&   12.42&       4.54&   39.54&   38.84&37.14&    252&     50&        75&        7.14 \\
330.0&NLRG&FR\,2&   &0.550& 3168&11.08&   11.29&$<$11.54&$<$11.62&$\sim$ 0.83&$<$37.86&$<$38.20&36.87&$<$  10&$<$  21&$\sim$  40&$\sim$  7.57 \\
332.0&BLRG&FR\,2&   &0.152&  716&10.87&   10.70&   10.78&$<$10.80&$\sim$ 0.93&   37.14&$<$37.38&35.05&    122&$<$ 211&$\sim$  40&$\sim$  6.75 \\
334.0& QSR&   L&   &0.555& 3203&11.54&$<$12.03&   12.20&   11.68&       3.30&   38.58&   38.11&36.47&    127&     44&        75&        6.39 \\
336.0& QSR&   L&   &0.927& 6036&11.79&   11.93&$<$12.64&$<$12.75&$\sim$ 0.77&$<$38.83&$<$39.32&37.09&$<$  55&$<$ 168&$\sim$  40&$\sim$  8.69 \\
&&&&& &&&&&&&&&&&&\\
337.0&NLRG&FR\,2&   &0.635& 3777&10.66&$<$11.60&$<$12.08&$<$12.04&$\sim$ 1.09&$<$38.36&$<$38.60&36.66&$<$  50&$<$  86&$\sim$  40&$\sim$  7.96 \\
340.0&NLRG&FR\,2&   &0.775& 4830&10.77&$<$11.75&$<$12.22&$<$12.37&$\sim$ 0.70&$<$38.51&$<$38.94&36.83&$<$  48&$<$ 131&$\sim$  40&$\sim$  8.32 \\
343.1&NLRG&FR\,2&CSS&0.750& 4638&11.13&$<$11.83&$<$12.46&$<$12.89&$\sim$ 0.37&$<$38.65&$<$39.48&36.84&$<$  64&$<$ 431&$\sim$  40&$\sim$  8.85 \\
351.0& QSR&   L&   &0.371& 1975&11.80&   12.22&   12.27&   11.87&       2.50&   38.68&   38.39&36.13&    360&    185&        50&        7.30 \\
352.0&NLRG&FR\,2&   &0.806& 5071&10.57&$<$11.56&$<$12.10&$<$12.27&$\sim$ 0.68&$<$38.37&$<$38.84&36.92&$<$  28&$<$  84&$\sim$  40&$\sim$  8.22 \\
&&&&& &&&&&&&&&&&&\\
356.0&NLRG&FR\,2&   &1.079& 7296&11.23&   11.52&   12.02&   12.25&       0.59&   38.28&   38.83&37.27&     10&     37&        40&        8.20 \\
368.0&NLRG&FR\,2&   &1.131& 7738&11.35&$<$12.39&   12.92&$<$12.82&$\sim$ 1.26&   39.21&$<$39.36&37.41&     63&$<$  89&$\sim$  40&$\sim$  8.73 \\
371.0& BL Lac&     &&0.051&  224&10.77&$<$10.70&$<$10.63&$<$10.23&$\sim$ 2.49&$<$37.10&$<$36.73&33.59&$<$3227&$<$1373&$\sim$  40&$\sim$  6.11 \\
380.0& QSR&   C&CSS&0.692& 4199&11.88&$<$12.28&$<$12.33&$<$11.85&$\sim$ 3.01&$<$38.76&$<$38.31&37.46&$<$  20&$<$   7&$\sim$  65&$\sim$  6.81 \\
381.0\parbox{0cm}{$^{\rm x}$}&NLRG&FR\,2&   &0.160&  758&10.43&   10.91&   11.10&   10.65&       2.82&   37.51&   37.15&35.34&    147&     64&        40&        6.70 \\
&&&&& &&&&&&&&&&&&\\
390.3&BLRG&FR\,2&   &0.056&  246&10.30&   10.64&   10.65&   10.23&       2.61&   37.13&   36.76&34.80&    211&     90&        40&        6.13 \\
401.0&LERG&FR\,2&   &0.201&  976&10.72&$<$10.15&$<$10.51&$<$11.00&$\sim$ 0.33&$<$36.52&$<$37.56&35.67&$<$   7&$<$  78&$\sim$  40&$\sim$  6.94 \\
405.0&NLRG&FR\,2&   &0.056&  246& 9.95&   10.55&   11.38&   11.30&       1.19&   37.64&   37.84&37.07&      4&      6&        50&        6.78 \\
427.1&LERG&FR\,2&   &0.572& 3324&10.59&$<$10.90&$<$11.99&$<$12.29&$\sim$ 0.50&$<$37.92&$<$38.85&36.89&$<$  11&$<$  89&$\sim$  40&$\sim$  8.21 \\
437.0&NLRG&FR\,2&   &1.480&10827&11.15&$<$12.35&$<$13.33&$<$13.12&$\sim$ 1.63&$<$39.50&$<$39.60&37.76&$<$  55&$<$  70&$\sim$  60&$\sim$  8.21 \\
&&&&& &&&&&&&&&&&&\\
441.0&NLRG&FR\,2&   &0.707& 4311&10.72&$<$11.72&$<$12.26&$<$12.42&$\sim$ 0.68&$<$38.52&$<$38.99&36.81&$<$  51&$<$ 150&$\sim$  40&$\sim$  8.36 \\
445.0&BLRG&FR\,2&   &0.056&  247&10.26&   10.87&   10.98&   10.50&       3.02&   37.46&   37.03&34.52&    855&    323&        50&        5.96 \\
454.3& FSQ&   C&OVV&0.860& 5498&12.50&$<$12.92&   13.53&   12.79&       5.59&   39.81&   39.13&37.06&    566&    118&       100&        7.10 \\
459.0&NLRG&FR\,2&   &0.220& 1081&10.89&   11.07&   11.83&   12.24&       0.39&   37.93&   38.81&35.85&    120&    908&        40&        8.18 \\
460.0&NLRG&FR\,2&   &0.268& 1353&10.39&$<$10.15&$<$10.43&$<$10.98&$\sim$ 0.28&$<$36.55&$<$37.58&35.56&$<$  10&$<$ 104&$\sim$  40&$\sim$  6.94 \\
&&&&& &&&&&&&&&&&&\\
\hline
\end{tabular}
\end{center} 
$^{\rm x}$ 3C381 could be classified as BLRG based on a noisy H$_\beta$ line,
which appears to be  broader than the [OIII]$\lambda$5007 line, but with considerable uncertainty (Saunders et al. 1989).
A new spectrum obtained at the KECK telescope shows that 3C381 should
be classified as NLRG (van Bemmel, private communication).
\end{table*}

\begin{table}
\caption[]{ Average values of log (R$_{\rm dr}$) for galaxies and quasars shown in
  Figs.\,\ref{msxxxx_fig_p80_to_p178_vs_z} and
  \ref{msxxxx_fig_p80_to_p178_vs_p178_ul}.
  The means $\pm$ standard deviations are given for two bins in
  luminosity of L$_{\rm IR}$ and P$_{\rm 178 MHz}$.  N denotes the
  number of sources in the bins.
  The averages are also calculated excluding extreme outliers.
\label{msxxxx_table_rdr}
}
\begin{center}
\footnotesize
\begin{tabular}{l|cr|cr}
bins                             &  quasars            &N       &    galaxies           &N     \\
\hline
Detections:		         &                     &        &                       &      \\
L$_{\rm IR    }$                 &                     &        &                       &      \\
$<$ 2$\cdot$10$^{12}$ L$_\odot$  &  1.946 $\pm$  0.290 & 6      &  1.907 $\pm$  0.620   &11    \\
			         &                     &        &  1.900 $\pm$  0.375   & 9\parbox{0cm}{$^a$}\\
$>$ 2$\cdot$10$^{12}$ L$_\odot$  &  1.818 $\pm$  0.327 &12      &  1.859 $\pm$  0.650   & 6    \\
			         &                     &        &  1.640 $\pm$  0.408   & 5\parbox{0cm}{$^b$}\\ 
\hline
Detections:		         &                     &        &                       &      \\
P$_{\rm 178 MHz}$                &                     &        &                       &      \\
$<$ 2$\cdot$10$^{28}$ W/Hz       & 1.970 $\pm$  0.301  & 8      &  2.174 $\pm$  0.548   &10    \\
                                 &                     &        &  1.960 $\pm$  0.351   & 8\parbox{0cm}{$^c$}    \\
$>$ 2$\cdot$10$^{28}$ W/Hz       & 1.774 $\pm$  0.308  &10      &  1.484 $\pm$  0.466   & 7    \\
                                 &                     &        &  1.603 $\pm$  0.376   & 6\parbox{0cm}{$^d$}    \\
\hline
Upper limits:		         &                     &        &                       &      \\
P$_{\rm 178 MHz}$                &                     &        &                       &      \\
$<$ 2$\cdot$10$^{28}$ W/Hz       & 2.197 $\pm$  0.533  &  7     &  2.117 $\pm$  0.382   & 8    \\
                                 & 2.040 $\pm$  0.367  &  6\parbox{0cm}{$^e$} &                       &      \\
$>$ 2$\cdot$10$^{28}$ W/Hz       & 1.819 $\pm$  0.434  &  8     &  1.935 $\pm$  0.337   &12    \\
                                 & 1.957 $\pm$  0.203  &  7\parbox{0cm}{$^f$} &                       &      \\
\hline
\end{tabular}
\end{center}
$^a$ excluding the extremes 3C\,321 and 3C405\\
$^b$ excluding the extreme 3C\,459\\
$^c$ excluding the extremes 3C\,321 and 3C459\\
$^d$ excluding the extreme  3C\,405\\
$^e$ excluding the extreme 3C\,371 (BL Lac type, outside Figure ranges)\\
$^f$ excluding the extreme 3C\,380 \\
\end{table}
\acknowledgements 
We thank the anonymous referee for constructive suggestions and
for encouraging us to discuss the receding torus model. 
The ISOPHOT Data Centre at MPIA is supported by 
Deutsches Zentrum f\"ur Luft- und Raumfahrt e.V. (DLR) with funds of 
Bundesministerium f\"ur Bildung 
und Forschung, grant no. 50\,QI\,0201. 
IRAM is supported by INSU/CNRS (France), MPG (Germany) and IGN (Spain).
Martin Haas thanks for grants from the Nordrhein-Westf\"alische Akademie der Wissenschaften,
funded by the Federal State Nordrhein-Westfalen and the Federal Republic of Germany.
This research is essentially based on data from the ISO and JCMT/SCUBA Archives.  
For literature and photometry search NED and SIMBAD were used.
It is a pleasure to thank IRAM for discretionary observing time 
with the 30\,m telescope at Pico Veleta.

\end{document}